\pdfoutput=1 
\documentclass[10pt,journal,twocolumn,letterpaper]{IEEEtran}
\usepackage{graphicx} 
\usepackage{listings}
\usepackage{xcolor}
\usepackage{tcolorbox}
\usepackage{afterpage}
\usepackage{soul}
\usepackage{hyperref}
\usepackage{graphicx}
\usepackage{wrapfig}
\usepackage{amsmath} 
\usepackage{fancyhdr}

\title{Understanding Graph Databases: A Comprehensive Tutorial and Survey}

\date{November 2024}

\author{
    \IEEEauthorblockN{
        Sydney Anuyah\textsuperscript{1}, 
        Emmanuel Bolade\textsuperscript{2}, 
         Oluwatosin Agbaakin\textsuperscript{1}, 
    }\\
    \IEEEauthorblockA{\textsuperscript{1}Luddy School of Informatics, Computing and Engineering, Indiana University, Indianapolis, IN, USA \\
    \textsuperscript{2}Data Science Department, Edyah Consulting, Anthony, Lagos, Nigeria \\
    Emails: sanuyah@iu.edu, vbolade@edyahlimited.com.ng, olbaagba@iu.edu}

}

\begin{document}
\maketitle
\section*{Abstract}
This tutorial is curated as a one-stop shop for understanding graph databases, as it emphasizes the foundations of graph theory and explores practical applications across multiple fields. The paper begins with foundational concepts, and further explains the structure of graphs through nodes and edges, including various types such as undirected, directed, weighted, and unweighted graphs. The tutorial outlines essential graph properties, terminologies, and key algorithms for network analysis, including Dijkstra’s shortest path algorithm and techniques for calculating node centrality and graph connectivity.

Graph databases, as discussed in the tutorial, offer advantages over traditional relational databases by enabling efficient management of complex, interconnected data. Overall, this paper discusses prominent graph database systems like Neo4j, Amazon Neptune, and ArangoDB, each with unique features for handling large datasets. There are practical instructions on implementing graph operations using NetworkX and Neo4j cover node and edge creation, attribute assignment, and advanced queries with Cypher. Furthermore, the paper includes common graph visualization techniques using tools such as Plotly and Neo4j Bloom which enhance the usability of graph data, while also focusing on community detection algorithms, such as the Louvain method, which support clustering in large networks. We conclude by providing future directions to researchers who are looking to enter into the world of graphs.

\vspace{1cm}
\noindent\textbf{Keywords:} Graph databases, Neo4j, NetworkX, graph theory, social network analysis, graph visualization, data modeling, graph algorithms, Cypher queries, community detection, graph neural networks, database management systems.

\section{Introduction}
\subsection{Understanding Graph Theory}
\subsubsection{What are Graphs? Nodes and Edges} 
The simplest definition of  a graph is a diagram with interconnected point. At least, that was how it was presented in mathematics. In this context, graphs are defined as abstract data structures consisting of nodes (or vertices) connected by edges. Graphs are  now are widely used to represent relationships between entities, and they are the go-to structure for training large language models\cite{yang2024give}. A graph is therefore an ideal way to model networks in various fields. These fields include geosciences, brain studies, transportation networks, etc.In geosciences, \cite{schrodt2020graph} nodes can represent geographic locations, while edges capture physical or spatial relationships between locations; For networked systems such as transportation systems, \cite{sadavare2012review} nodes can represent cities or transport hubs, while the edges represents the routes or pipelines connecting them, etc. Such graph representations provide a comprehensive view of the structure of a network, allowing exploration of connections and dependencies \cite{sadavare2012review, schrodt2020graph}.

In biomedical research, graphs can be used to model brain connectivity, as \cite{falsaperla2021graph} and \cite{ismail2020graph} illustrated in their studies of pediatric epilepsy and brain connectivity. Nodes here represented the distinct brain regions and the edges indicated functional connections observed through EEG data.  The grpah structure aided researchers to analyze the influence of specific brain areas on cognitive functions. Graph analysis in these cases ususally enables the identification of key nodes critical to neural activity \cite{falsaperla2021graph, ismail2020graph}

\subsubsection{Types of Graphs: Undirected, Directed, Weighted, Unweighted} 
Graphs come in several types based on the properties of their edges, including undirected, directed, weighted, and unweighted graphs. Undirected graphs have edges with no particular direction, making them suitable for representing mutual relationships like social connections, as \cite{sanchez2021gentle} illustrate in their study of social networks. Here, friendship or mutual interactions are best represented through undirected graphs where both nodes are equally connected \cite{sanchez2021gentle}. In contrast, directed graphs involve edges with a defined direction, as discussed by \cite{sadavare2012review} in the context of supply chain networks, where goods flow sequentially from one location to another, requiring a directed approach.

Graphs may also be weighted, where edges are assigned weights that reflect the strength or cost of the connection. In transportation networks, weights can represent distances or time costs, optimizing the routing and minimizing travel expenses \cite{sadavare2012review}. Unweighted graphs, on the other hand, treat all connections equally and are commonly used when only the existence of a relationship is relevant, as seen in brain connectivity research, where edges merely indicate connectivity without weighing them \cite{ismail2020graph}. Understanding which graph type to apply depends on the requirements of the application, as improper usage can lead to misinterpretation of the data \cite{glazer2011challenges}. 

\subsubsection{Basic Terminology and Notations}
Graph theory uses specific terminologies and notations that form its foundation. Concepts like degree, path, and adjacency matrix are crucial to analyze connectivity within graphs. The degree of a node, which indicates the number of edges connected to it, is essential in spatial network studies. \cite{schrodt2020graph} explain that degree is used in geosciences to measure a node’s influence within a spatial network, assisting in identifying regions of high connectivity and impact. Similarly, in studies on graph complexity, nodes with high degrees contribute significantly to a network’s structural intricacy, as described by \cite{kim2008complex}, particularly in biological and social systems where interconnectedness can signify greater complexity \cite{kim2008complex}; \cite{schrodt2020graph}.

Other key terms, such as paths and adjacency matrices, aid in computational modeling. Paths represent routes or sequences of edges connecting nodes, which are fundamental in transportation systems where shortest path algorithms, like Dijkstra’s and Bellman-Ford, are used to optimize network flows \cite{sadavare2012review}. An adjacency matrix, a square matrix that represents graph connections, simplifies visualization and manipulation of complex networks, as highlighted in EEG-based brain network analysis for exploring connectivity patterns. \cite{falsaperla2021graph}. 


%

\subsection{Real-World Applications of Graphs}
\subsubsection{Social Networks} 
Social networks are among the most common applications of graph theory, allowing for the modeling and analysis of complex relationships between users. Graph theory enables social network analysis (SNA) through metrics that evaluate network structure, user influence, and connectivity \cite{majeed2020graph}. For example, link prediction algorithms anticipate potential connections, suggesting friends or followers based on existing network patterns \cite{daud2020applications}. These models capture essential social interactions and enable applications in community detection, anomaly detection, and influence analysis. Furthermore, recent advancements in graph neural networks (GNNs) have enhanced the scalability and precision of social network analysis, enabling more sophisticated user recommendations and deeper insights into network dynamics \cite{bhatti2023deep, anuyah2024can}. Through these methodologies, social networks leverage graph theory to foster user engagement, enhance information dissemination, and improve security.

\subsubsection{Transportation Networks} 
Graph theory also plays a crucial role in transportation networks by modeling routes and optimizing logistics. Each location, such as a city or transit hub, is represented as a node, while the connections between these locations (e.g., roads or railways) serve as edges. Shortest path algorithms like Dijkstra's and Bellman-Ford are fundamental in reducing travel time and costs across transportation networks \cite{sadavare2012review}. Additionally, graph neural networks (GNNs) have recently been applied to traffic forecasting, leveraging data such as road congestion and passenger flow to predict traffic patterns accurately \cite{jiang2022graph}. These predictions allow for real-time adjustments in transportation services, benefiting urban planners and logistics companies. The application of GNNs to transportation networks has revolutionized traffic management, enhancing efficiency and improving commuter experiences.

\subsubsection{Recommender Systems} 
In recommender systems, graphs enable the modeling of user-item interactions, forming the foundation for personalized recommendations. Users and items are represented as nodes, with edges connecting users to the items they have interacted with or rated. Recent advancements in GNNs have significantly enhanced the efficacy of recommender systems by utilizing the high-order connectivity inherent in graph data \cite{wu2022graph}. These models consider not only direct connections but also the broader network structure, capturing indirect relationships that improve recommendation accuracy \cite{gao2022graph}. This approach enables platforms like e-commerce sites and streaming services to provide tailored suggestions, reducing information overload for users and improving their overall experience. Graph-based recommendation techniques have become essential in handling the vast data of modern recommender systems, making them more relevant and efficient.

\subsubsection{Biological Networks} 
In biological research, graph theory is instrumental in analyzing the interactions within complex biological systems, such as protein-protein interactions or gene regulatory networks. Nodes represent bioentities like proteins or genes, while edges denote the interactions between them, enabling insights into biological processes and disease mechanisms \cite{koutrouli2020guide}. The increase in biological data has led to the application of deep learning techniques, especially GNNs, to handle and analyze complex biological networks. GNNs are particularly useful in predicting protein functions, drug discovery, and understanding genetic interactions, where they provide a high level of precision in detecting patterns within heterogeneous biological data \cite{muzio2021biological, jin2021application}. By applying graph theory to biological networks, researchers can better understand disease pathways and identify potential therapeutic targets, contributing significantly to advancements in bioinformatics and medicine.

As graph theory continues to evolve, its methodologies allow researchers to tackle increasingly complex, data-rich environments, as evidenced by its wide adoption in fields such as neuroscience, urban planning, and social science research \cite{schrodt2020graph, velivckovic2023everything}. Graph theory’s capacity to model both tangible and abstract relationships has made it essential for analyzing connectivity and interactions in modern data science.

\subsection{Graph Databases}
Graph databases have emerged as specialized systems designed to handle data with complex, interconnected relationships. Unlike traditional relational databases that store data in tables, graph databases represent data as nodes (entities) and edges (relationships), making them highly effective for applications where connections between data points are central. These databases support advanced analytics, pathfinding, and pattern matching, and are instrumental in fields such as social networking, recommendation systems, and fraud detection \cite{besta2023demystifying, jabdelearning}. As large datasets become increasingly common, graph databases offer efficient solutions for managing and querying complex data structures with minimal latency and high scalability. 

\subsubsection{Differences between Graph Databases and Relational Databases}
Relational databases (RDBMS) and graph databases differ significantly in terms of data structure, storage, and query mechanisms. Relational databases organize data in tables, enforcing data integrity through primary and foreign keys. While effective for structured, transaction-based data, RDBMS struggle with queries involving multi-level relationships, as JOIN operations become costly with complex queries \cite{deshmukhreview, jabdelearning}. Graph databases, in contrast, use nodes to represent entities and edges to capture relationships, creating a natural and efficient structure for traversing relationships without JOINs \cite{sahu2020ubiquity}. This makes graph databases particularly suitable for applications requiring deep relationship exploration, such as recommendation engines and social networks \cite{besta2023demystifying, angles2020mapping}.

\subsubsection{Graph Databases: Neo4j} 
Neo4j is one of the most widely used graph databases, recognized for its robust property graph model and flexibility in managing and traversing complex relationships. Neo4j organizes data using a labeled property graph model, where both nodes and edges can have attributes stored as key-value pairs. This model allows Neo4j to efficiently perform graph traversals, a crucial feature for applications like fraud detection, social network analysis, and recommendation engines \cite{rabuzin2022supporting, besta2023demystifying}. Additionally, Neo4j uses the Cypher query language, specifically designed for querying graph structures. Cypher’s syntax is intuitive, making it accessible for both developers and data scientists, and allows for complex pattern matching and relationship-based queries that are challenging in relational databases \cite{angles2020mapping}.
Neo4j is also known for its high level of consistency and support for ACID (Atomicity, Consistency, Isolation, Durability) transactions, which are essential for data reliability. As Neo4j supports both transactional and analytical operations, it is particularly favorable for applications requiring real-time analytics. For example, in social networks, Neo4j can help identify influential nodes (individuals) and track interactions across a network, providing insights into social dynamics \cite{majeed2020graph}. Neo4j’s scalability is also noteworthy; it can be configured in a clustered environment for horizontal scaling, making it suitable for handling massive datasets commonly seen in enterprise applications \cite{jabdelearning}. 

\subsubsection{Graph Databases: Amazon Neptune} 
Amazon Neptune is a managed graph database service by AWS, supporting both RDF and Property Graph models. Its compatibility with both SPARQL and Gremlin query languages allows users to choose between querying RDF-based semantic data or traditional graph data, making it versatile for diverse applications. Neptune’s architecture is designed for high availability and fault tolerance, with automated backups and replication across multiple availability Zones, ensuring minimal downtime and data reliability \cite{atemezing2021empirical}. One of Neptune’s key features is its seamless integration with other AWS services, which simplifies data ingestion, processing, and analytics on the AWS cloud \cite{atemezing2021empirical}.

Neptune is particularly effective for applications that require fast query response times and efficient handling of large-scale, dynamic data. Its use cases include recommendation engines, fraud detection, and knowledge graph applications, where rapid data retrieval and complex relationship analysis are essential \cite{besta2023demystifying}. For instance, in a recommendation system, Neptune can process high-order connections by analyzing the relationships between users, products, and behaviors, resulting in highly personalized recommendations. Additionally, Neptune’s RDF support makes it valuable for semantic data processing, allowing organizations to model and query domain-specific ontologies, making it especially popular in knowledge graph development and metadata management \cite{cheng2021multi}.

One limitation of Neptune, however, is its dependency on the AWS ecosystem, which may restrict flexibility for users who rely on multi-cloud strategies. Nonetheless, for enterprises already invested in AWS, Neptune offers a powerful, scalable, and secure solution for managing complex data relationships in real-time \cite{atemezing2021empirical}. 

\subsubsection{Graph Databases: ArangoDB} 
ArangoDB is a unique graph database known for its multi-model capabilities, supporting document, key-value, and graph data models within a single database. This flexibility allows users to handle diverse data types and structures, which is beneficial for applications with complex and varied data needs. ArangoDB uses the Arango Query Language (AQL), which is designed to work across different data models, providing a unified query language that simplifies complex queries involving multiple data types \cite{belgundi2023analysis}. This versatility makes ArangoDB suitable for big data applications, as it can manage different data structures without needing multiple databases.

ArangoDB’s multi-model architecture allows it to support both OLTP (Online Transaction Processing) and OLAP (Online Analytical Processing) queries, enabling it to handle both transactional and analytical workloads. For instance, in e-commerce, ArangoDB can support a real-time recommendation system using its graph model while managing inventory data with its document model. This multi-functional capability is particularly useful in use cases like data integration and ETL (Extract, Transform, Load) processes, where data from various sources need to be combined and processed efficiently \cite{mavrogiorgos2021comparative}. Additionally, ArangoDB’s support for graph-based analytics enables it to handle complex tasks such as fraud detection, network analysis, and predictive modeling.

ArangoDB also offers excellent scalability, supporting horizontal scaling through sharding and distributed cluster setups. Its integration with Kubernetes enhances its adaptability in cloud environments, providing flexibility in deployment and management \cite{belgundi2023analysis}. This makes it a viable option for organizations handling large datasets with complex requirements, such as those in finance, healthcare, and retail industries. 

\subsubsection{Benefits of Using Graph Databases}
Graph databases provide several advantages over traditional relational databases, especially when managing complex, interconnected data. One primary benefit is their ability to efficiently model relationships directly in the data structure, which eliminates the need for complex JOIN operations. This structure is especially advantageous in applications requiring deep relationship analysis, such as social networks, where it is essential to traverse and analyze complex user interactions rapidly \cite{majeed2020graph, koshy2023data, sahu2020ubiquity}. Graph databases also offer schema flexibility, allowing for dynamic adjustments to data structures without the need for costly schema migrations, a valuable feature for evolving datasets \cite{jabdelearning}

Another significant benefit of graph databases is their scalability. Databases like Amazon Neptune and ArangoDB are designed to handle large datasets while maintaining query efficiency, making them suitable for enterprise-level applications. This scalability is critical in big data environments, where traditional relational databases may struggle with performance issues as data volume and complexity increase \cite{atemezing2021empirical, belgundi2023analysis}. Additionally, graph databases support advanced analytics, such as community detection, link prediction, and centrality measures, which are essential for applications in areas like recommendation systems, fraud detection, and biological network analysis \cite{cheng2021multi, muzio2021biological}.

\section{Creating and Visualizing Basic Graphs}
In this section, we’ll discuss how to create basic graphs using NetworkX in Python and Neo4j, focusing on creating nodes and edges, assigning attributes, and loading and saving data in Neo4j. These operations form the foundation for building and visualizing more complex graph structures.

\subsection{Creating Nodes and Edges}
Creating nodes and edges is the foundational step in building a graph. Nodes represent entities, and edges represent the relationships or interactions between these entities.

In NetworkX: NetworkX is a Python library that simplifies the creation and manipulation of graph structures. Nodes and edges can be added directly using simple commands.

Creating Nodes: You can create individual nodes or add multiple nodes at once.

Here is how to create nodes in Python

\begin{lstlisting}[language=Python, caption={Creating Nodes in Python with NetworkX}]
import networkx as nx

G = nx.Graph()  # Initializes an undirected graph

# Adding nodes individually
G.add_node("Alice")
G.add_node("Bob")

# Adding multiple nodes
G.add_nodes_from(["Charlie", "David", "Eve"])

\end{lstlisting}

\begin{lstlisting}[language=Python, caption={Viewing created Nodes in Python}]
# View the nodes in the graph
print("Nodes in the graph:", G.nodes())
\end{lstlisting}

\begin{tcolorbox}[colback=gray!10, colframe=gray!80!black, title= Nodes]
Nodes in the graph: ['Alice', 'Bob', 'Charlie', 'David', 'Eve']
\end{tcolorbox}

\begin{figure}[h!]
    \centering
    \includegraphics[width=0.5\textwidth]{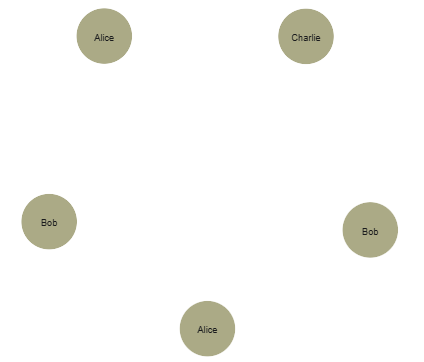} 
    \caption{Neo4j visualization of nodes and edges}
    \label{fig:neo4j_nodes_edges}
\end{figure}

Creating Edges: Edges can be added similarly, either one at a time or in groups. An edge defines a connection between two nodes.
\begin{lstlisting}[language=Python, caption={Creating Edges in Python with NetworkX}]
# Adding an edge between two nodes
G.add_edge("Alice", "Bob")

# Adding multiple edges
G.add_edges_from([("Alice", "Charlie"), ("Bob", "David"), ("Charlie", "Eve")])

#Print out the edges for a view
print("Edges in the graph:", G.edges())
\end{lstlisting}

\begin{tcolorbox}[colback=gray!10, colframe=gray!80!black, title= Edges]
Edges in the graph: [('Alice', 'Bob'), ('Alice', 'Charlie'), ('Bob', 'David'), ('Charlie', 'Eve')]
\end{tcolorbox}

\vspace{1cm}

\begin{figure}[h!]
    \centering
    \includegraphics[width=0.5\textwidth]{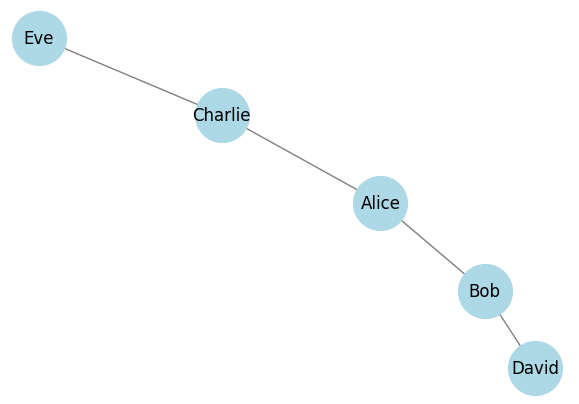} 
    \caption{Graph visualization of edges created in Python with NetworkX}
    \label{fig:graph_edges_example}
\end{figure}

The resulting graph visualization is shown in Figure~\ref{fig:graph_edges_example}.

In Neo4j, nodes and relationships (edges) are created using the Cypher query language, which is designed for handling graph data. Neo4j allows you to specify types for both nodes and relationships, making it suitable for more structured and complex data.

\begin{lstlisting}[language=, caption={Creating Nodes with Neo4j}]
CREATE (a:Person {name: 'Alice'})
CREATE (b:Person {name: 'Bob'})
\end{lstlisting}

In the commands above, the labels \texttt{Person} identify the nodes as representing people, and each node has a property \texttt{name}. \\

Creating Relationships (Edges) in Neo4j

\begin{lstlisting}[language=, caption={Creating Edges with Neo4j}]
MATCH (a:Person {name: 'Alice'}), (b:Person {name: 'Bob'})
CREATE (a)-[:FRIEND]->(b)
\end{lstlisting}

This command creates a \texttt{FRIEND} relationship between Alice and Bob, indicating a connection in the context of a social network \cite{jabdelearning, rabuzin2022supporting}. \\

\textit{Assigning Attributes to Nodes and Edges} \\
Attributes can provide additional information about nodes and edges, such as a person’s age or a relationship’s strength.

In NetworkX: Attributes can be assigned when nodes or edges are created or added later.

\begin{lstlisting}[language=Python, caption={Assigning Attributes to Nodes in NetworkX}]
# Adding a node with attributes
G.add_node("Alice", age=30, city="New York")

# Adding attributes to an existing node
G.nodes["Bob"]["age"] = 25
G.nodes["Bob"]["city"] = "Los Angeles"
\end{lstlisting} 

Viewing created Node-Attributes in Python
\begin{lstlisting}[language=Python, caption={Viewing created Node-Attributes in Python}]
#To view the attributes of a specific node we use
print("The attributes for the node Bob are: ", G.nodes["Bob"])
#To view the attributes of multiple nodes we use
print("The attributes for the node Bob and Alice are: ", {node: G.nodes[node] for node in ["Bob", "Alice"]})
#To view the attributes of all nodes we use
print("The attributes for all the nodes are: ", G.nodes(data=True))
\end{lstlisting} 
\vspace{1cm}

\begin{tcolorbox}[colback=gray!10, colframe=gray!80!black, title= Printing the Attributes of the Nodes]
The attributes for the node Bob are:  {'age': 25, 'city': 'Los Angeles'}\\
\\
The attributes for the node Bob and Alice are:  {'Bob': {'age': 25, 'city': 'Los Angeles'}, 'Alice': {'age': 30, 'city': 'New York'}}\\
\\
The attributes for all the nodes are:  [('Alice', {'age': 30, 'city': 'New York'}), ('Bob', {'age': 25, 'city': 'Los Angeles'}), ('Charlie', {}), ('David', {}), ('Eve', {})]
\end{tcolorbox}

\begin{figure}[h!]
    \centering
    \includegraphics[width=0.5\textwidth]{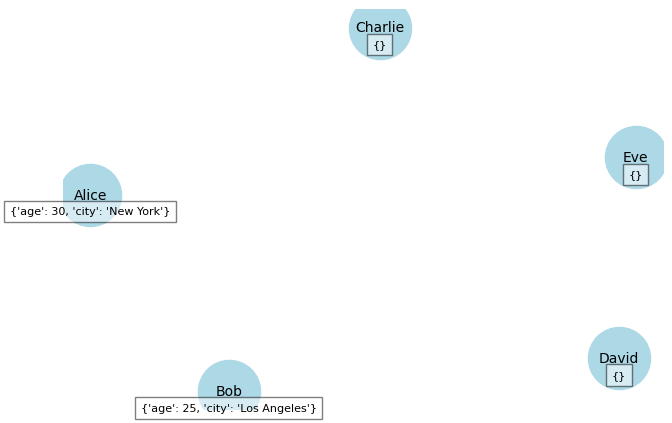}
    \caption{Graph visualization of nodes with attributes in NetworkX}
    \label{fig:node_attributes_example}
\end{figure}

The visualization of nodes with their attributes is shown in Figure~\ref{fig:node_attributes_example}.

Note: In NetworkX, you can dynamically add or modify both nodes and attributes for every node and edge even if they were not predefined initially like the relationship "weight" and the node "Faith" below. 
\vspace{1cm}

\begin{lstlisting}[language=Python, caption={Assigning Attributes to Edges}]
# Adding an attribute and defining the weight class + Defining a new node
G.add_edge("Alice", "Faith", relationship="friends", weight=4)

# Adding an attribute to an existing edge
G.edges["Alice", "Charlie"]["weight"] = 3
\end{lstlisting} 
\vspace{1cm}

\begin{lstlisting}[language=Python, caption={Viewing created Edge-Attributes in Python}]
# To view the attributes of a specific edge
print("The attributes for the edge between Alice and Bob are:", G.edges["Alice", "Bob"])

# To view the attributes of multiple specific edges
print("The attributes for the edges between Alice-Bob and Alice-Charlie are:", 
      {edge: G.edges[edge] for edge in [("Alice", "Bob"), ("Alice", "Charlie")]})

# To view the attributes of all edges
print("The attributes for all the edges are:", G.edges(data=True))
\end{lstlisting} \vspace{1cm}

\begin{tcolorbox}[colback=gray!10, colframe=gray!80!black, title= Printing the Attributes of the Edges]
The attributes for the edge between Alice and Bob are: {'relationship': 'friends', 'weight': 4} \\
\\
The attributes for the edges between Alice-Bob and Alice-Charlie are: {('Alice', 'Bob'): {'relationship': 'friends', 'weight': 4}, ('Alice', 'Charlie'): {'weight': 3}} \\
\\
The attributes for all the edges are: [('Alice', 'Bob', {'relationship': 'friends', 'weight': 4}), ('Alice', 'Charlie', {'weight': 3}), ('Alice', 'Faith', {'relationship': 'friends', 'weight': 4}), ('Bob', 'David', {}), ('Charlie', 'Eve', {})]
\\
\end{tcolorbox}

\begin{figure}[h!]
    \centering
    \includegraphics[width=0.5\textwidth]{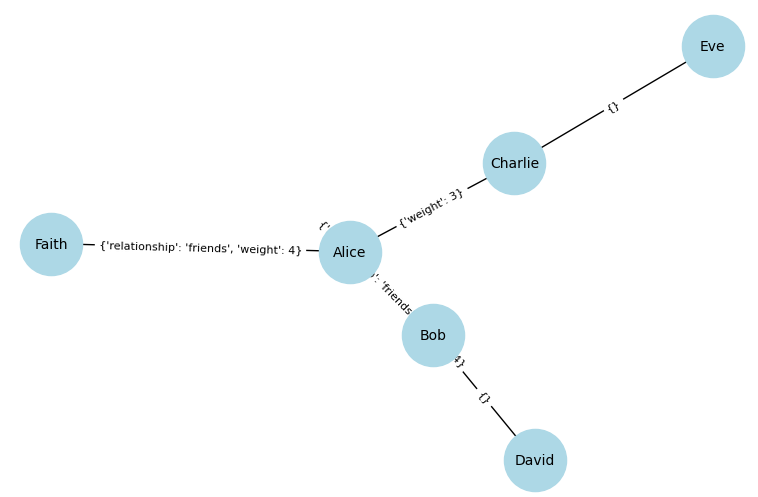}
    \caption{Graph visualization of edges with attributes in NetworkX}
    \label{fig:edge_attributes_example}
\end{figure}

The visualization of edges with their attributes is shown in Figure~\ref{fig:edge_attributes_example}.

In Neo4j, attributes (called properties in Neo4j) are assigned using Cypher when creating nodes and relationships.
\\
\begin{lstlisting}[language=, caption={Assigning Attributes to Nodes in Neo4j}]
CREATE (a:Person {name: 'Alice', age: 30, city: 'New York'})
\end{lstlisting}

\vspace{1cm}
\begin{lstlisting}[language=, caption={Assigning Attributes to Relationships in Neo4j}]
MATCH (a:Person {name: 'Alice'}), (b:Person {name: 'Bob'})
CREATE (a)-[:FRIEND {since: 2015, closeness: 4}]->(b)

\end{lstlisting}
In this example, the \texttt{FRIEND} relationship between Alice and Bob has properties \texttt{since} and \texttt{closeness}, adding contextual information to the connection. Assigning attributes enhances the graph’s informational depth, allowing for more meaningful queries and analyses, especially in applications where metadata plays a crucial role \cite{deshmukhreview}. \\

Loading and saving graph data are crucial operations in Neo4j, particularly when working with large datasets. Neo4j provides several methods for data import and export, including CSV import for batch processing and APOC procedures for more complex tasks. Neo4j supports importing CSV files, which is especially useful for large datasets. Data can be loaded using the \texttt{LOAD CSV} command in Cypher.

\vspace{1cm}

\begin{lstlisting}[language=, caption={Loading Data from a CSV File}]
LOAD CSV WITH HEADERS FROM 'file:///people.csv' AS row
CREATE (p:Person {name: row.name, age: toInteger(row.age), city: row.city})
\end{lstlisting}

This command reads a CSV file \texttt{(people.csv)} and creates a \texttt{Person} node for each row, with attributes assigned from the CSV columns. Neo4j’s \texttt{LOAD CSV} command is powerful, allowing large datasets to be processed efficiently \cite{rabuzin2022supporting}. Neo4j allows data to be exported through the APOC (Awesome Procedures on Cypher) library, which provides extensive functionality for managing data. Nodes, relationships, and entire graphs can be exported to formats like CSV, JSON, or even Cypher queries.

\vspace{1cm}
\begin{lstlisting}[language=, caption={Saving Data in Neo4j}]
CALL apoc.export.csv.all('exported_graph.csv', {useTypes: true, delimiter: ';'})
\end{lstlisting}
This code exports the entire database to a CSV file. The \texttt{useTypes} option ensures that node and relationship types are preserved, while the \\texttt{delimiter} specifies the separator to be used in the file. This functionality is especially useful for creating backups or transferring data to other systems \cite{jabdelearning, besta2023demystifying}

\subsection{Using Neo4j in Python}
To work with Neo4j in Python, we use the Neo4j Python Driver to connect to a Neo4j database and execute Cypher queries. This allows Python-based applications to interact directly with Neo4j. Before starting, ensure that the Neo4j Python driver is installed.

\begin{lstlisting}[language=Python, caption={Install Neo4j}]
pip install neo4j
\end{lstlisting}

Next, connect to a running Neo4j database instance. For local setups, the default URL is typically \href{bolt://localhost:7687}{bolt://localhost:7687}, and you will need to provide the database credentials. To create the Neo4j Database, you can do the following:

\begin{itemize}
    \item Go to the \href{https://neo4j.com/download/}{Neo4j Download Page} and download Neo4j Desktop. This is a standalone app that provides an easy way to install and manage local Neo4j databases. Follow the installation instructions for your operating system. Once installed, open Neo4j Desktop.
    \item In Neo4j Desktop, create a new project to organize your databases. You can name it anything you would like. For example, “MyFirstGraphDatabase.” Note that the names should not have spaces, or it will show an error.
    \item Within the project, create a new database by clicking “Add Database” and choosing “Local DBMS.” Set a name for the database (e.g., \texttt{test\_database}) and a password of choice. Note: The default username will be \texttt{neo4j}. Start the database by clicking the play button.
    \item After starting the database, click on it to see the connection details. You will see the bolt URL (\href{bolt://localhost:7687}{bolt://localhost:7687}), necessary for connecting to Neo4j from Python. If using Chrome, typing in \href{localhost:7687}{localhost:7687} may also work.
\end{itemize}

Now, connect to the Neo4j Database from Python. Use the following template code, replacing "your\_password" with the password you set when creating the database.
\begin{lstlisting}[language=Python, caption={Connect to Neo4j Database from Python}]
uri = "bolt://localhost:7687"
username = "neo4j"
password = "your_password"  # Replace with your password
driver = GraphDatabase.driver(uri, auth=(username, password))
print("Connected to Neo4j!")
\end{lstlisting}

\begin{tcolorbox}[colback=gray!10, colframe=gray!80!black, title=Output message]
Connected to Neo4j!
\end{tcolorbox}

Now, let's run Cypher queries to interact with the database.

\begin{lstlisting}[language=Python, caption={Creating Nodes using Neo4j from Python}]
def create_simple_nodes(tx):
    tx.run("CREATE (a:Person {name: 'Alice'})")
    tx.run("CREATE (b:Person {name: 'Bob'})")

# Run the transaction
with driver.session() as session:
    session.write_transaction(create_simple_nodes)
    print("Created nodes Alice and Bob.")
\end{lstlisting}


After this, confirm that the nodes have been created by running the following command:

\begin{lstlisting}[language=Python, caption={Retrieving the Nodes using Neo4j from Python}]
# Function to run the MATCH query and retrieve all Person nodes
def get_person_nodes(tx):
    result = tx.run("MATCH (n:Person) RETURN n")
    for record in result:
        print(record["n"])  # Prints each Person node
# Run the code within a session
with driver.session() as session:
    print("Person nodes in the database:")
    session.read_transaction(get_person_nodes)
\end{lstlisting}

\begin{tcolorbox}[colback=gray!10, colframe=gray!80!black, title=Output message]
Person nodes in the database: \\

Node element id='4:bd31ae9c-f686-4122-b53f-f8d6cb58adb9:0' labels=frozenset({'Person'}) properties={'name': 'Alice'}  
\\
Node element id='4:bd31ae9c-f686-4122-b53f-f8d6cb58adb9:1' labels=frozenset({'Person'}) properties={'name': 'Bob'} 
\end{tcolorbox}

To get cleaner output, change the query as follows:

\begin{lstlisting}[language=Python, caption={Retrieving Cleaned Nodes Output}]
def get_person_nodes(tx):
    result = tx.run("MATCH (n:Person) RETURN n.name AS name")
    for record in result:
        name = record["name"]
        print(f"Name: {name}")
with driver.session() as session:
    print("Person nodes in the database:")
    session.read_transaction(get_person_nodes)
\end{lstlisting}

\begin{tcolorbox}[colback=gray!10, colframe=gray!80!black, title=Output message]
Person nodes in the database:
Name: Alice
Name: Bob
\end{tcolorbox}

\begin{figure}[h]
    \centering
    \begin{minipage}{0.15\textwidth}
        \centering
        \includegraphics[width=\textwidth]{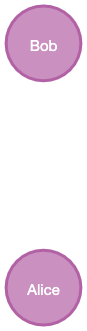}
        \caption{Nodes created in Neo4j}
        \label{fig:image1}
    \end{minipage}\hfill
    \begin{minipage}{0.15\textwidth}
        \centering
        \includegraphics[width=\textwidth]{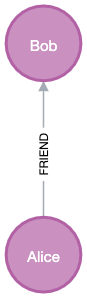}
        \caption{Relationships created in Neo4j}
        \label{fig:image2}
    \end{minipage}
\end{figure}

Figure \ref{fig:image1} shows the graph output in Neo4j. Next, we define relationships as follows:

\begin{lstlisting}[language=Python, caption={Creating Edges using Neo4j from Python}]
def create_friend_relationship(tx):
    tx.run("""
        MATCH (a:Person {name: 'Alice'}), (b:Person {name: 'Bob'})
        CREATE (a)-[:FRIEND]->(b)
    """)

# Run the transaction
with driver.session() as session:
    session.write_transaction(create_friend_relationship)
    print("Created FRIEND relationship between Alice and Bob.")
\end{lstlisting}

\begin{tcolorbox}[colback=gray!10, colframe=gray!80!black, title=Output message]
Created FRIEND relationship between Alice and Bob as seen in Figure \ref{fig:image2}
\end{tcolorbox}

\subsection{Case Study: Simple Social Network Graph}

\subsubsection{Objective}
This case study demonstrates the creation of a simple social network graph, where nodes represent individuals and edges define friendships. Each person has attributes such as \textit{age} and \textit{city}, and each friendship has a \textit{closeness} level to indicate relationship strength.

\subsubsection{Scenario}
The social network consists of five individuals: Alice, Bob, Charlie, David, and Eve. Each person lives in a different city and has varying ages. Friendships are represented by edges, each with a unique closeness level. This network structure helps analyze relationships within social groups and identify key connections.

\subsubsection{Graph Creation in Python with NetworkX}
To create this network, we use NetworkX, a Python library for graph-based data modeling.

\begin{lstlisting}[language=Python, caption={Creating a Simple Social Network Graph with NetworkX}]
import networkx as nx
import matplotlib.pyplot as plt

# Create a graph instance
G = nx.Graph()

# Add nodes with attributes
G.add_node("Alice", age=30, city="New York")
G.add_node("Bob", age=25, city="Los Angeles")
G.add_node("Charlie", age=35, city="Chicago")
G.add_node("David", age=28, city="San Francisco")
G.add_node("Eve", age=32, city="Boston")

# Add edges with a 'closeness' attribute
G.add_edge("Alice", "Bob", closeness=5)
G.add_edge("Alice", "Charlie", closeness=4)
G.add_edge("Bob", "David", closeness=3)
G.add_edge("Charlie", "Eve", closeness=4)
G.add_edge("David", "Eve", closeness=2)

# Draw the graph with node and edge attributes
plt.figure(figsize=(10, 8))
pos = nx.spring_layout(G)
nx.draw(G, pos, with_labels=True, node_color='lightblue', node_size=3000, font_size=10)

# Display node attributes
node_labels = {node: f"{node}\nAge: {G.nodes[node]['age']}, City: {G.nodes[node]['city']}" for node in G.nodes}
nx.draw_networkx_labels(G, pos, labels=node_labels, font_size=8)

# Display edge attributes
edge_labels = {(u, v): f"Closeness: {G.edges[u, v]['closeness']}" for u, v in G.edges}
nx.draw_networkx_edge_labels(G, pos, edge_labels=edge_labels, font_size=8)

# Save and display the graph
plt.savefig("social_network_example.png", format="png")
plt.show()
\end{lstlisting}

\subsubsection{Visualization}
\begin{figure*}[h!]
    \centering
    \includegraphics[width=\textwidth]{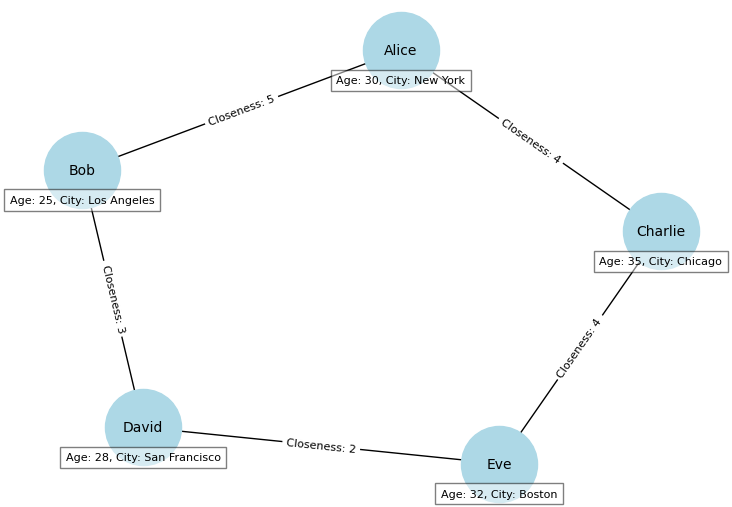}
    \caption{Visualization of the Simple Social Network Graph}
    \label{fig:social_network_example}
\end{figure*}

Figure~\ref{fig:social_network_example} displays each individual as a node with age and city information, while friendship closeness is represented on edges.

This example demonstrates how to construct a social network graph and visualize it effectively with attributes on both nodes and edges. Such graphs are useful in social network analysis, where relationships and individual characteristics are essential for understanding network structures.

\section{Graph Properties and Basic Operations}
\subsection{Definitions}
\subsubsection{Degree}
The degree of a node (or vertex) in a graph is defined as the number of connections or edges it has, which is a fundamental indicator of the node’s connectivity within the graph. This property is crucial for assessing various metrics, such as centrality and influence, in social network analysis. Specifically, degree centrality is helpful in identifying influential nodes, as highlighted by  \cite{selvi2023square}.
Moreover, general degree distance measures further capture aspects of connectivity and network efficiency, as discussed by \cite{vetrik2021general}. Beyond social networks, the degree metric holds significant value in network security contexts. For instance, understanding the degree of vertices can help in evaluating network vulnerabilities \cite{alshaer2018identifying} and determining vertex importance in complex networked systems, as shown in studies on betweenness and other centrality measures by \cite{giustolisi2019b}.

\subsubsection{Path Length}
Path length refers to the shortest path or distance between two nodes, representing reachability within the network. This metric is critical in routing and communication applications. Distance measures based on path lengths, such as shortest and detour paths, help quantify network efficiency \cite{mustafa2021m_n}. Like diametral paths, path-based analyses have applications in understanding complex network topologies, where analyzing shortest paths improves routing and resource allocation \cite{mangam2017diametral}.

\subsubsection{Adjacency Matrix}
An adjacency matrix represents the connections between nodes in a graph, where each matrix element indicates the presence or absence of an edge between node pairs. This matrix format is essential for visualizing and calculating various graph properties, such as similarity and distances, as explored by  \cite{gervens2022graph}.
The adjacency matrix serves as a foundational tool in defining the structural properties of a graph and is particularly valuable in specialized fields like protein network analysis, as noted by \cite{aguilar2021catalytic}. Its utility extends to computational applications, enabling efficient matrix-based algorithms for graph traversals and spectral analysis \cite{dratman2022rank}.
In structural analyses, adjacency matrices represent and manipulate graph data. For instance, they allow for the calculation of spectral radii, a measure that reflects the connectivity strength within a network \cite{wills2020metrics}.

\subsubsection{Distance Measures}
Distance measures, such as shortest paths, play a crucial role in determining network connectivity and resilience. These metrics help map communication patterns within networks and enhance our understanding of network dynamics, as explored by \cite{noschese2024network}. Additionally, distance measures are often used in conjunction with degree properties to assess structural differentiation in specialized applications, such as brain functional networks \cite{li2020maximum}.
Distance measures quantify the separation between nodes, offering valuable insights into network structure and layout. Common metrics include the shortest path distance, alongside specialized measures like the Fréchet distance, which is used for embedded graphs \cite{akitaya2021distance}. Recent research has also introduced the reciprocal degree distance (RDD), a robust metric for evaluating vertex connectivity and network resilience \cite{an2019reciprocal}.

\subsection{Pathfinding and Shortest Path Algorithms}
\subsubsection{Dijkstra's Algorithm}
Dijkstra's algorithm calculates the shortest path from a source node to all other nodes in a graph, with a time complexity of O((m + n) logn), where m is the number of edges and n is the number of nodes. This algorithm is effective for networks where edge weights are non-negative, as it iteratively selects the shortest known path until the destination node is reached or all nodes have been visited. Comparisons with A* have shown that while Dijkstra is generally more resource-intensive for complex environments, it remains preferable in grids without heuristic guidance \cite{vceslis2020efficiency}.

\subsubsection{Implementation in NetworkX}
NetworkX, a Python library for graph manipulation, provides an efficient way to apply Dijkstra’s algorithm. 

\begin{lstlisting}[language=Python, caption={Retrieving Edges using Neo4j from Python}]
import networkx as nx

# Create a graph and add weighted edges
G = nx.Graph()
G.add_weighted_edges_from([(1, 2, 7), (1, 3, 9), (2, 4, 10), (3, 4, 2)])

# Compute shortest path using Dijkstra's algorithm
shortest_path = nx.dijkstra_path(G, source=1, target=4)
path_length = nx.dijkstra_path_length(G, source=1, target=4)

print("Shortest path:", shortest_path)
print("Path length:", path_length)
\end{lstlisting} 

\begin{tcolorbox}[colback=gray!10, colframe=gray!80!black, title=Output message]
Shortest path: [1, 3, 4] \\
Path length: 11
\end{tcolorbox}

\subsubsection {Implementation in Neo4j}
Neo4j, a graph database, provides the \textit{`gds.shortestPath'} function to implement Dijkstra’s algorithm efficiently in large datasets. 

\begin{lstlisting}[language=Python, caption={Identifying the shortest route using Dijkstra's algorithm }]
MATCH (start:Location {name: "A"}), (end:Location {name: "B"})
CALL gds.shortestPath.dijkstra.stream({
  sourceNode: start,
  targetNode: end,
  relationshipWeightProperty: 'distance'
})
YIELD nodeId, cost
RETURN gds.util.asNode(nodeId).name AS node, cost
\end{lstlisting} 

\begin{tcolorbox}[colback=gray!10, colframe=gray!80!black, title=Output message]
Shortest path from A to B:
Node: A, Cumulative Cost: 0 \\
Node: C, Cumulative Cost: 5 \\
Node: B, Cumulative Cost: 10 \\
\end{tcolorbox}

This query identifies the shortest route between two locations based on a weighted property (e.g., distance), suitable for applications in logistics and supply chain management where cost minimization is important \cite{keung2022shortest}.
\subsubsection {Use Cases and Applications}
\begin{itemize}
\item Financial Networks: Dijkstra’s algorithm aids in portfolio management by minimizing transaction costs in financial asset graphs, where finding 
cost-efficient paths improves rebalancing 
strategies in trading 

\cite{vallarino2024dynamic}.
\item Gaming and Simulation: Dijkstra's algorithm is used in game development to calculate optimal paths in grid-based environments, offering precise pathfinding that supports AI decision-making for navigating virtual spaces \cite{kuushalie2024efficiency}.
\item Robotic Navigation: In robotic fulfillment systems, Dijkstra’s algorithm is foundational for path planning in warehouses, optimizing routes within complex layouts to improve operational efficiency \cite{keung2022shortest}.
\end{itemize}

\subsubsection {Algorithmic Efficiency Studies}
Studies highlight Dijkstra's efficiency, especially in unweighted or lightly weighted networks, though adaptations like parallel computing or combining with other algorithms (e.g., GNNs) can further enhance performance. For instance, combining pathfinding with Graph Neural Networks (GNNs) improves predictions and reduces transaction costs in finance \cite{cappart2023combinatorial}.

\subsection{Graph Density and Connectivity Analysis}

Graph density measures the number of edges in a graph compared to the maximum possible number of edges. For an undirected graph \( G \) with \( n \) nodes and \( m \) edges, the density \( D \) can be calculated by:
\[
D = \frac{2m}{n(n-1)}
\]
In directed graphs, the formula is adjusted based on directed edges. High graph density indicates a well-connected graph, which is often essential in applications requiring robust connectivity, like transportation and social networks \cite{kerzner2017graffinity}.

\subsubsection{Connected Components}
Connected components represent maximal subgraphs in which any two nodes are connected by paths. Identifying these components is imperative for network segmentation and isolation tasks. For instance, in dynamic graphs, connected components can reveal network changes over time, making them suitable for tracking connectivity in evolving networks \cite{vernet2023study}.

\subsubsection{Implementation in Python (NetworkX)}
NetworkX offers tools to identify connected components:

\begin{lstlisting}[language=Python, caption={Identifying connected components in a graph}]
import networkx as nx

G = nx.Graph([(1, 2), (2, 3), (4, 5)])
components = list(nx.connected_components(G))
print("Connected Components:", components)
\end{lstlisting}

Here, nodes are grouped into distinct connected components, which is useful for graph clustering and community detection.

\begin{tcolorbox}[colback=gray!10, colframe=gray!80!black, title=Output message]
Connected Components: [{1, 2, 3}, {4, 5}]
\end{tcolorbox}

\begin{figure}[h!]
    \centering
    \includegraphics[width=0.5\textwidth]{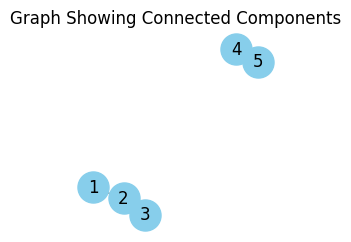}
    \caption{Graph Showing Two Connected Components: \{1, 2, 3\} and \{4, 5\}}
\end{figure}

\subsubsection{Connectivity Measures}
Connectivity measures quantify a graph's robustness to disconnections. Key metrics include vertex connectivity (the minimum number of nodes that need removal to disconnect the graph) and edge connectivity (the minimum number of edges that must be removed). These measures assess network resilience, especially in critical infrastructure and fault-tolerant systems \cite{jafarpour2020building}.

\subsubsection{Advanced Connectivity Techniques}
Recent research explores extended connectivity properties like algebraic connectivity, calculated from the second-smallest eigenvalue of the graph Laplacian matrix, providing insights into graph stability and connectedness \cite{xue2019algebraic}. Incremental algorithms also dynamically track connectivity changes in real time for adaptive systems, which is particularly valuable in network monitoring 
\cite{georgiadis2018incremental}.

\subsection{Basic Graph Operations for Graph Analysis}

\subsubsection{Node and Edge Creation}
In Neo4j, defining nodes and edges enables graph construction and connectivity modeling, foundational for resilience studies. For example, creating a logistics network might look like:
\begin{lstlisting}
CREATE (n1:Node {name: 'Warehouse'})
CREATE (n2:Node {name: 'Distribution Center'})
CREATE (n1)-[:CONNECTS {distance: 50}]->(n2)
\end{lstlisting}
This structure aids in modeling and assessing critical links within networks, such as supply chains or urban infrastructure \cite{sun2020performance}.

\subsubsection{Shortest Path for Resilience}
Shortest path algorithms in Cypher are key for understanding redundancy and finding alternate paths. In Neo4j, this can be done using:

\begin{lstlisting}[language=Python, caption={Finding the shortest path between two nodes using Cypher}]
MATCH (start:Node {name: 'A'}), (end:Node {name: 'B'})
MATCH path = shortestPath((start)-[*]-(end))
RETURN path;
\end{lstlisting}

This query helps identify alternative routes within networks, essential for recovery in disrupted networks, such as transport or electric grids.

\begin{tcolorbox}[colback=gray!10, colframe=gray!80!black, title=Output message]
Shortest path from A to B:
Node: A \\
Node: C \\
Node: B \\
\end{tcolorbox}

\begin{figure}[h!]
    \centering
    \includegraphics[width=0.5\textwidth]{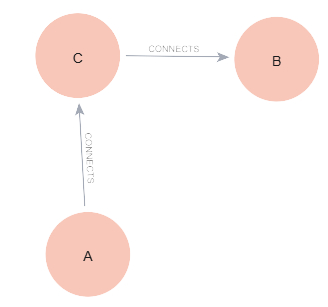}
    \caption{Graph Visualization of the Shortest Path from A to B}
\end{figure}

This query helps identify alternative routes within networks, essential for recovery in disrupted networks, such as transport or electric grids \cite{mirjalili2023resilience}.

\subsubsection{Node and Edge Deletion for Fault Tolerance}
Assessing network resilience often involves simulating node or edge failures. By removing connections in Cypher, analysts can evaluate a network’s robustness:

\begin{lstlisting}[language=Python, caption={Removing an edge between two nodes}]
MATCH (n1:Node {name: 'A'})-[r:CONNECTS]->(n2:Node {name: 'B'})
DELETE r;
\end{lstlisting}

This operation tests how network structure changes affect connectivity, providing insights into fault tolerance.

\begin{tcolorbox}[colback=gray!10, colframe=gray!80!black, title=Output message]
Connection between Node A and Node B has been removed.
\end{tcolorbox}

\begin{figure}[h!]
    \centering
    \includegraphics[width=0.45\textwidth]{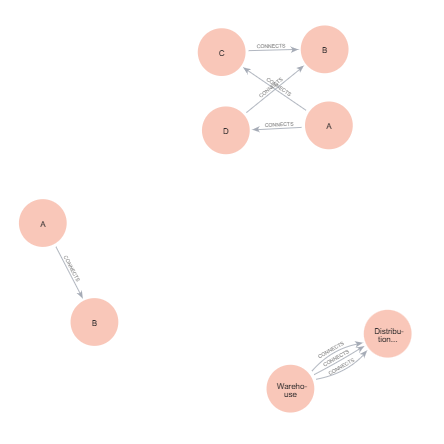}
    \caption{Graph Visualization Before Edge Deletion}
\end{figure}

\begin{figure}[h!]
    \centering
    \includegraphics[width=0.45\textwidth]{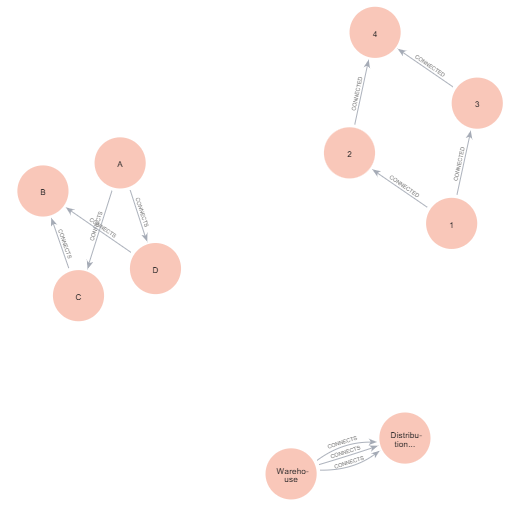}
    \caption{Graph Visualization After Edge Deletion (Node A to Node B removed)}
\end{figure}

This operation tests how network structure changes affect connectivity, providing insights into fault tolerance \cite{yuge2024operational}.

\subsubsection{Subgraph Analysis for Core Resilience}
Neo4j allows for subgraph extraction, useful for examining connected components or community resilience. In scenarios where critical nodes are removed, this helps identify isolated subgraphs:
\begin{lstlisting}
CALL gds.wcc.stream({
  nodeProjection: 'Node',
  relationshipProjection: 'CONNECTS'
})
YIELD componentId, nodeId
RETURN componentId, gds.util.asNode(nodeId).name AS nodeName
\end{lstlisting}
This analysis identifies weak points within a network, highlighting nodes that increase resilience when protected \cite{laishram2018measuring}.

\subsubsection{Case Study: Network Resilience in a Transportation System}
A practical application of these operations can be seen in resilience analysis of transportation networks, where resilience measures are vital for post-disaster recovery. Studies on New York City’s transportation resilience during snowstorms utilized graph metrics to estimate operational continuity with reduced data, which can be replicated in Neo4j for real-time disruption analysis \cite{mirjalili2023resilience}.

\section{Manipulating Graphs in NetworkX and Neo4j}

\subsection{Adding and Removing Nodes and Edges}

\subsubsection{Adding Nodes and Edges}
Adding nodes and edges is essential for growing or updating networks, which is common in dynamic environments such as social networks, infrastructure models, and IoT systems.

\paragraph{In NetworkX:}
NetworkX provides straightforward methods for adding single or multiple nodes and edges:
\begin{lstlisting}
import networkx as nx
G = nx.Graph()

# Adding a single node
G.add_node("A")

# Adding multiple nodes
G.add_nodes_from(["B", "C", "D"])

# Adding a single edge
G.add_edge("A", "B")

# Adding multiple edges
G.add_edges_from([("A", "C"), ("B", "D")])
\end{lstlisting}
This code enables quick network expansion. NetworkX also supports adding attributes to nodes and edges, which is useful for enriching nodes with properties like "weight" or "type" for specific applications.

\paragraph{In Neo4j:}
In Neo4j, nodes and edges (referred to as "relationships") can be added using Cypher commands:
\begin{lstlisting}
// Adding a node with properties
CREATE (a:Person {name: 'Alice', age: 30})
// Adding an edge (relationship) between nodes
MATCH (a:Person {name: 'Alice'}), (b:Person {name: 'Bob'})
CREATE (a)-[:KNOWS]->(b)
\end{lstlisting}
Adding nodes and edges with properties allows for flexible data modeling in Neo4j, where entities are enriched with contextual data, enabling deeper insights in subsequent analysis. Studies on network expansion techniques, such as plug-and-play approaches, indicate that adding nodes should prioritize maintaining connectivity to avoid fragmentation, which is critical for network stability \cite{studli2021plug}.

\subsubsection{Removing Nodes and Edges}
Removing elements from a network is often required for scenarios such as simulating node failures, pruning unnecessary connections, or analyzing the effects of disruptions.

\paragraph{In NetworkX:}
Removing nodes and edges in NetworkX is efficient, supporting single and batch removals:
\begin{lstlisting}
# Removing a single node
G.remove_node("A")

# Removing multiple nodes
G.remove_nodes_from(["B", "C"])

# Removing a single edge
G.remove_edge("A", "B")

# Removing multiple edges
G.remove_edges_from([("A", "C"), ("B", "D")])
\end{lstlisting}
NetworkX’s API ensures that any attempt to remove non-existent nodes or edges raises an error, helping maintain data integrity.

\paragraph{In Neo4j:}
Cypher supports selective deletion, allowing precise control over node and relationship removal. When removing a node, the associated relationships must also be removed:
\begin{lstlisting}
// Remove a node and all its relationships
MATCH (n:Person {name: 'Alice'})
DETACH DELETE n
\end{lstlisting}
This command removes both the node and any connecting relationships, ensuring no orphaned relationships remain, which is important for data integrity. In large-scale networks, especially IoT and social networks, strategies for intelligent edge and node removal optimize robustness against failures. The Intelligent Rewiring (INTR) mechanism, for example, enhances resilience by redistributing connectivity in scale-free networks \cite{abbas2022towards}.

\subsubsection{Modifying Nodes and Edges}
Modifying the properties of nodes and edges is essential for dynamically updating network attributes, enhancing edge attributes for pathfinding, reliability, or transport optimization.

\paragraph{In NetworkX:}
NetworkX supports adding or updating attributes for nodes and edges, which is useful for applications like traffic modeling or infrastructure analysis:
\begin{lstlisting}
# Adding or modifying a node attribute
G.nodes["A"]["type"] = "hub"

# Adding or modifying an edge attribute
G["A"]["B"]["weight"] = 4
\end{lstlisting}
These modifications allow the network to evolve, adjusting characteristics such as edge weights based on real-time data or node roles based on updated contexts.

\paragraph{In Neo4j:}
Cypher’s \texttt{SET} command enables efficient updates to node and edge properties, supporting incremental changes in large datasets:
\begin{lstlisting}
// Modify a node property
MATCH (a:Person {name: 'Alice'})
SET a.age = 31

// Modify an edge property
MATCH (a)-[r:KNOWS]->(b)
SET r.since = 2021
\end{lstlisting}
This flexibility in Cypher supports operations like updating relationships to model evolving social connections or transactional changes in networks.

In performance-focused applications, modifying edge weights or properties to optimize transport efficiency or connectivity in real-time is crucial. Studies on rewiring strategies suggest that targeted edge modifications based on centrality measures (e.g., betweenness) can substantially enhance network capacity and maintain structural stability in scale-free networks \cite{kumari2020efficient}.

\subsubsection{Practical Applications and Efficiency}

Efficient manipulation of nodes and edges is essential for large networks, such as in IoT, infrastructure modeling, and social network analysis. Neo4j’s indexing and query optimization make it well-suited for handling high volumes of relationship data, while NetworkX’s graph manipulation methods are effective for network simulations and analyses requiring flexibility and fast prototyping.

\begin{itemize}
    \item \textbf{Neo4j}: For applications requiring real-time analysis, Neo4j’s Cypher queries enable fast lookups and modifications, making it ideal for continuously evolving networks where each update can impact overall resilience or connectivity.
    \item \textbf{NetworkX}: NetworkX’s Python-based functions provide the flexibility to experiment with graph changes, simulate network effects, and validate different manipulation strategies.
\end{itemize}

\subsection{Working with Subgraphs}

\subsubsection{Extracting Subgraphs}

Extracting subgraphs is a common operation in graph analysis, allowing for focused exploration of specific regions or patterns within a large graph.

\begin{itemize}
    \item \textbf{In NetworkX}: NetworkX provides tools for creating subgraphs based on node or edge criteria. For instance, to extract a subgraph that includes only nodes connected to a particular central node, we can use the following code:
    \begin{lstlisting}[language=Python, caption={Extracting a subgraph of nodes adjacent to node 1 in NetworkX}]
import networkx as nx
G = nx.Graph()
G.add_edges_from([(1, 2), (2, 3), (3, 4), (4, 5), (1, 5)])

# Extracting a subgraph containing nodes adjacent to node 1
nodes = list(G.neighbors(1)) + [1]  # Nodes adjacent to 1 plus node 1 itself
subG = G.subgraph(nodes)

# Display nodes and edges in the subgraph
print("Nodes in subgraph:", subG.nodes())
print("Edges in subgraph:", subG.edges())
    \end{lstlisting}

    This creates a subgraph containing all nodes adjacent to node 1, providing a smaller, focused view of the network.

    \begin{tcolorbox}[colback=gray!10, colframe=gray!80!black, title=Expected Output]
    Nodes in subgraph: [1, 2, 5] \\
    Edges in subgraph: [(1, 2), (1, 5)]
    \end{tcolorbox}

    \item \textbf{In Neo4j}: In Neo4j, subgraph extraction often involves Cypher queries that filter nodes and relationships based on specified criteria. For example, to extract a subgraph that includes only nodes with specific properties, use:

\begin{lstlisting}[language=, caption={Extracting a subgraph of people over 30 and their relationships in Neo4j}]
MATCH (n:Person)-[r:KNOWS]-(m:Person)
WHERE n.age > 30
RETURN n, r, m
    \end{lstlisting}

    This query extracts a subgraph of people over the age of 30 and their relationships, isolating a subset of the graph based on property filters.

    \begin{tcolorbox}[colback=gray!10, colframe=gray!80!black, title=Expected Output]
    Nodes with $age > 30$ and their relationships are returned as part of the subgraph: \\
    
    Example:
    \begin{tabular}{|c|c|c|}
    \hline
    Node 1 & Relationship & Node 2 \\
    \hline
    Alice & KNOWS & Bob \\
    Charlie & KNOWS & Dana \\
    \hline
    \end{tabular}
    \end{tcolorbox}

\end{itemize}

\subsubsection{Filtering Techniques for Subgraph Extraction}

Filtering techniques are important for managing the scope and size of subgraphs, especially in large networks where analyzing the entire graph is computationally prohibitive.

\begin{itemize}
    \item \textbf{Vertex-based Filtering}: This technique filters nodes based on attributes or connectivity. In a social network, for example, filtering by nodes with high centrality scores can provide a subgraph of the most influential nodes.
    
    \textbf{In NetworkX}:
    \begin{lstlisting}[language=Python, caption={Extracting a subgraph of nodes with degree 2 or higher}]
# Subgraph with nodes of degree 2 or more
high_degree_nodes = [n for n, d in G.degree() if d >= 2]
subG = G.subgraph(high_degree_nodes)

# Display nodes and edges in the filtered subgraph
print("Nodes in filtered subgraph:", subG.nodes())
print("Edges in filtered subgraph:", subG.edges())
    \end{lstlisting}

    \begin{tcolorbox}[colback=gray!10, colframe=gray!80!black, title=Expected Output]
    Nodes in filtered subgraph: [1, 2, 3] \\
    Edges in filtered subgraph: [(1, 2), (2, 3)]
    \end{tcolorbox}

    \item \textbf{Edge-based Filtering}: This technique filters edges based on weights or relationship types. In transport networks, filtering by high-weight (traffic-heavy) edges can reveal critical paths or bottlenecks.

    \textbf{In NetworkX}:
    \begin{lstlisting}[language=Python, caption={Extracting a subgraph with edges above a certain weight threshold}]
# Subgraph with edges having a weight above a certain threshold
heavy_edges = [(u, v) for u, v, d in G.edges(data=True) if d.get("weight", 0) > 5]
subG = G.edge_subgraph(heavy_edges)

# Display nodes and edges in the filtered subgraph
print("Nodes in filtered subgraph:", subG.nodes())
print("Edges in filtered subgraph:", subG.edges())
    \end{lstlisting}

    \begin{tcolorbox}[colback=gray!10, colframe=gray!80!black, title=Expected Output]
    Nodes in filtered subgraph: [2, 3, 4] \\
    Edges in filtered subgraph: [(2, 3), (3, 4)]
    \end{tcolorbox}

    \item \textbf{Sampling-based Filtering}: Techniques like random node sampling or feature-driven sampling reduce the graph size while preserving important features. These techniques are efficient for creating representative subgraphs from large graphs \cite{yang2023extract}.
    
    \textbf{In NetworkX (Example of Random Sampling)}:
    \begin{lstlisting}[language=Python, caption={Random sampling of nodes to create a representative subgraph}]
import random
sampled_nodes = random.sample(G.nodes(), 3)
subG = G.subgraph(sampled_nodes)

# Display nodes and edges in the sampled subgraph
print("Nodes in sampled subgraph:", subG.nodes())
print("Edges in sampled subgraph:", subG.edges())
    \end{lstlisting}

    \begin{tcolorbox}[colback=gray!10, colframe=gray!80!black, title=Expected Output]
    Nodes in sampled subgraph: [1, 3, 5] \\
    Edges in sampled subgraph: [(1, 3)]
    \end{tcolorbox}

\end{itemize}

\subsubsection{Advanced Subgraph Analysis Techniques}

Advanced methods, such as those used in subgraph matching and counting, help optimize the extraction of subgraph patterns that match specific structural criteria.

\begin{itemize}
    \item \textbf{Subgraph Matching}: Techniques like FaSTest reduce the sample space by filtering non-matching nodes or edges before analyzing potential subgraph matches, thereby improving efficiency \cite{shin2023cardinality}. These techniques are particularly valuable for tasks like network motif detection or pattern recognition in biological networks.

    \item \textbf{Multi-View Filtering}: Multi-view filters, as in the GMADL architecture, analyze subgraphs through multiple criteria or perspectives, enhancing the precision of subgraph extraction and aiding in classification tasks \cite{zheng2021subgraph}.

    Efficient subgraph extraction is vital for analyzing large graphs without excessive computational costs. Techniques like indexing, as highlighted by \cite{sun2019scaling}, enable fast retrieval of subgraphs by pre-indexing common structures, reducing the need for exhaustive searches.

    In large-scale graph analysis, bounding the size or complexity of subgraphs through filtering also optimizes performance. For instance, algorithms designed for specific girth constraints (the length of the shortest cycle) can quickly enumerate small subgraphs, improving both speed and accuracy in sparse graph structures \cite{kurita2018bounded}.
\end{itemize}

\subsection{Merging and Comparing Graphs}

\subsubsection{Union of Graphs}
The union of two graphs combines all nodes and edges from both graphs. In practical applications, union operations are useful for consolidating data from different graph sources, such as merging social networks or integrating infrastructure networks.

\begin{itemize}
    \item \textbf{In NetworkX}: NetworkX provides the \texttt{nx.compose()} function for union operations.
\end{itemize}
\begin{lstlisting}
import networkx as nx
G1 = nx.Graph()
G1.add_edges_from([(1, 2), (2, 3)])

G2 = nx.Graph()
G2.add_edges_from([(3, 4), (4, 5)])

# Union of G1 and G2
G_union = nx.compose(G1, G2)
\end{lstlisting}

This operation retains all unique nodes and edges from both graphs. For cases where graphs may have overlapping nodes with conflicting attributes, \text{$nx.compose_all()$} can handle multiple graphs and preserve their structure without overwriting attributes.

\begin{itemize}
    \item \textbf{In Neo4j}: In Neo4j’s Cypher, union operations can be achieved by merging nodes and relationships:
\end{itemize}
\begin{lstlisting}
MATCH (a:Person)-[:KNOWS]-(b:Person)
WITH collect(DISTINCT a) + collect(DISTINCT b) AS nodes
UNWIND nodes AS n
RETURN n
\end{lstlisting}

This example aggregates all unique nodes across different subgraphs, which can be useful for consolidating social networks or community structures. Graph unions are important in network science for tasks that require amalgamating graph structures to study composite properties or centralities. For instance, \cite{unnithan2019betweenness} explored betweenness centrality in composite graphs created by union operations, highlighting the significance of graph unions in network dynamics.

\subsubsection{Intersection of Graphs}
The intersection of two graphs includes only the nodes and edges common to both graphs. This operation is valuable for identifying shared connections, such as mutual friends in social networks or common pathways in biological networks.

\begin{itemize}
    \item \textbf{In NetworkX}: NetworkX supports intersection using \texttt{nx.intersection()}:
\end{itemize}
\begin{lstlisting}
# Intersection of G1 and G2
G_intersection = nx.intersection(G1, G2)
\end{lstlisting}

This function returns a new graph containing only the nodes and edges present in both \( G1 \) and \( G2 \), useful for focusing on core relationships or common elements between graphs.

\begin{itemize}
    \item \textbf{In Neo4j}: Cypher queries can filter overlapping nodes and edges:
\end{itemize}
\begin{lstlisting}
MATCH (a)-[r:FRIENDS]-(b)
WHERE (a)-[:COLLEAGUES]-(b)
RETURN a, b, r
\end{lstlisting}

This command extracts relationships that exist under both “FRIENDS” and “COLLEAGUES,” effectively creating an intersection based on relationship criteria. The intersection of graphs is essential in analyzing sub-networks or common structural elements in composite networks. For example, 

\cite{baumgartner2018term} demonstrated how graph intersections aid in anti-unification of term-graphs, focusing on bisimilar structures and common attributes.

\subsubsection{Difference of Graphs}
The difference operation between two graphs removes nodes and edges in one graph from another, often used to isolate unique or exclusive features.

\begin{itemize}
    \item \textbf{In NetworkX}: NetworkX provides \texttt{nx.difference()} to compute the difference:
\end{itemize}

\begin{lstlisting}[language=Python, caption={Computing the difference of two graphs in NetworkX}]
import networkx as nx
G1 = nx.Graph([(1, 2), (2, 3), (3, 4)])
G2 = nx.Graph([(2, 3)])

# Difference of G1 and G2
G_difference = nx.difference(G1, G2)

# Display nodes and edges in the difference graph
print("Nodes in difference graph:", G_difference.nodes())
print("Edges in difference graph:", G_difference.edges())
\end{lstlisting}

\begin{tcolorbox}[colback=gray!10, colframe=gray!80!black, title=Expected Output]
Nodes in difference graph: [1, 3, 4] \\
Edges in difference graph: [(1, 2), (3, 4)]
\end{tcolorbox}

This operation identifies unique edges in one graph that are absent in the other, such as unique connections in one part of a network or exclusive features in comparative studies.

\begin{itemize}
    \item \textbf{In Neo4j}: In Neo4j, a difference can be implemented by filtering nodes and relationships absent in the second graph:
\end{itemize}
\begin{lstlisting}
MATCH (a)-[r:KNOWS]-(b)
WHERE NOT (a)-[:COLLEAGUES]-(b)
RETURN a, r, b
\end{lstlisting}

This query extracts relationships labeled “KNOWS” that are not also labeled “COLLEAGUES,” isolating unique interactions in the “KNOWS” network. Graph differences help isolate exclusive relationships or features in comparative network studies, such as identifying unique interaction patterns in different community groups \cite{alhevaz2020generalized}.

\subsubsection{Comparing Graph Properties}
Comparing graphs involves analyzing their structural and spectral properties, often focusing on metrics like centrality, spectral properties, and graph distance measures.

\begin{itemize}
    \item \textbf{Spectral Analysis}: Spectral analysis involves studying the eigenvalues of graph adjacency or Laplacian matrices. For example, the join operation in graphs, which affects eigenvalues and distance spectra, plays a role in comparing network robustness and connectivity properties \cite{alhevaz2020generalized}.

    \item \textbf{Gromov-Wasserstein Distance}: The partial Gromov-Wasserstein distance, which maps nodes across partially overlapping graphs, facilitates complex comparison tasks where exact alignment is not feasible \cite{liu2022one}. This distance is effective in matching nodes with different structural roles, providing a distinct comparison of graph similarity.
\end{itemize}

\section{Advanced Graph Visualizations}

\subsection{Circular Layout}
The circular layout arranges nodes in a circular form, which is particularly effective for visualizing hierarchical data structures or networks where symmetry and clear adjacency relationships are crucial. Circular layouts make it easy to see clusters and cyclic dependencies in a graph, which can be beneficial in social network analysis or biological pathway visualization.

\subsubsection{Circular Layout of Nodes}
NetworkX provides the \texttt{nx.circular\_layout()} function to arrange nodes in a circle. This layout is useful for visualizing cyclic structures or ensuring equal spacing among nodes.

\begin{itemize}
    \item \textbf{In NetworkX}: To create a circular layout, we can use the following code:
\end{itemize}

\begin{lstlisting}[language=Python, caption={Arranging nodes in a circular layout using NetworkX}]
import networkx as nx
import matplotlib.pyplot as plt

G = nx.cycle_graph(10)  # Create a cycle graph with 10 nodes
pos = nx.circular_layout(G)  # Arrange nodes in a circle
nx.draw(G, pos, with_labels=True)  # Draw the graph
plt.show()  # Display the circular layout
\end{lstlisting}

\begin{figure}[h!]
    \centering
    \includegraphics[width=0.5\textwidth]{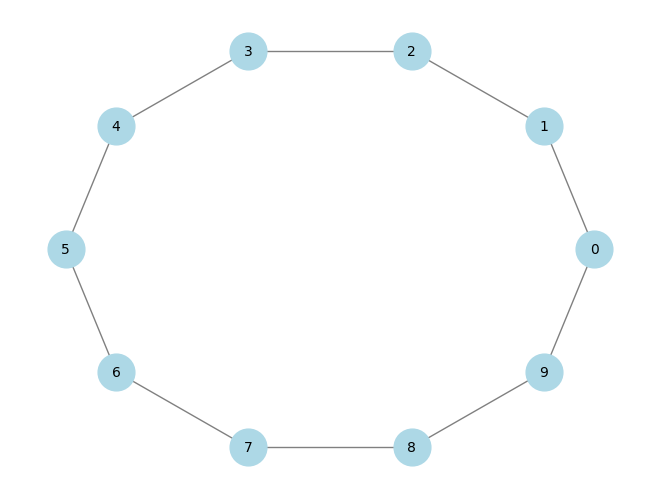}
    \caption{Circular Layout of a 10-Node Cycle Graph}
\end{figure}

This layout is useful for visualizing cycles or connected components, as each node is equally spaced, allowing clear visibility of connections and node degrees. In a study by \cite{kasyanov2021circular}, a circular layout was used in the Visual Graph system to visualize hierarchical graphs with attributed nodes, proving beneficial in fields like engineering and data science, where hierarchical relationships need representation.

\subsection{Spectral Layout}
Spectral layouts use eigenvectors of the graph’s Laplacian matrix to place nodes in a way that minimizes certain energy metrics. This layout is effective for organizing nodes in clustered structures, making it suitable for community detection or identifying dense areas within sparse graphs.

\begin{itemize}
    \item \textbf{In NetworkX}: NetworkX’s \texttt{nx.spectral\_layout()} function uses eigenvalues to determine node positions:
\end{itemize}

\begin{lstlisting}[language=Python, caption={Arranging nodes in a spectral layout using NetworkX}]
pos = nx.spectral_layout(G)
nx.draw(G, pos, with_labels=True)
plt.show()
\end{lstlisting}

\begin{figure}[h!]
    \centering
    \includegraphics[width=0.5\textwidth]{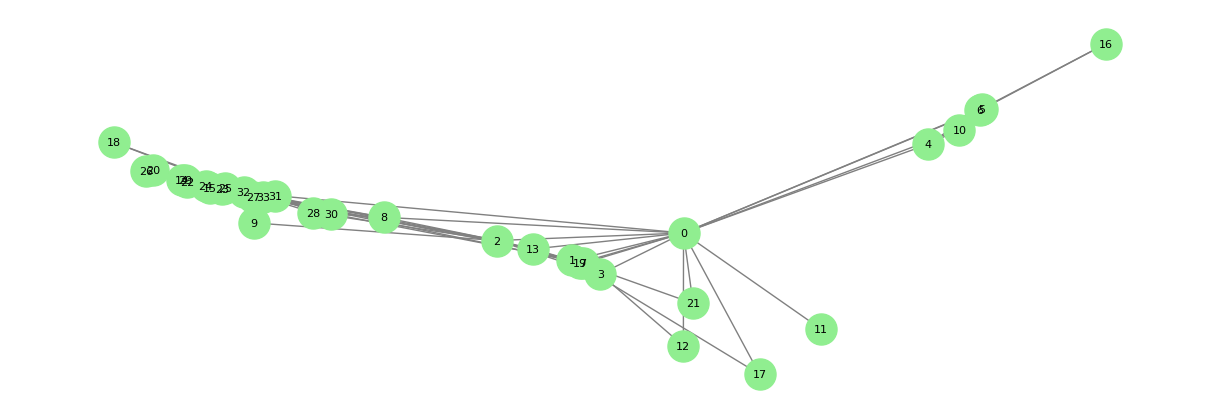}
    \caption{Spectral Layout of a Sample Graph}
\end{figure}

Spectral layouts can reveal underlying clusters or groupings within complex networks, making them valuable for analyzing community structures in social networks or functional modules in biological networks. Spectral properties effectively maintain node distances that reflect the graph's overall structure, supporting network clustering and optimization.

\subsection{Spring Layout (Force-Directed Layout)}
The spring layout, often implemented using the Fruchterman-Reingold algorithm, models the graph as a physical system where nodes repel each other while edges act like springs, pulling connected nodes together. This layout is commonly used for undirected graphs, balancing readability and symmetry.

\begin{itemize}
    \item \textbf{In NetworkX}: The spring layout in NetworkX is implemented via \texttt{nx.spring\_layout()}:
\end{itemize}

\begin{lstlisting}[language=Python, caption={Arranging nodes in a spring layout using NetworkX}]
pos = nx.spring_layout(G)
nx.draw(G, pos, with_labels=True)
plt.show()
\end{lstlisting}

\begin{figure}[h!]
    \centering
    \includegraphics[width=0.5\textwidth]{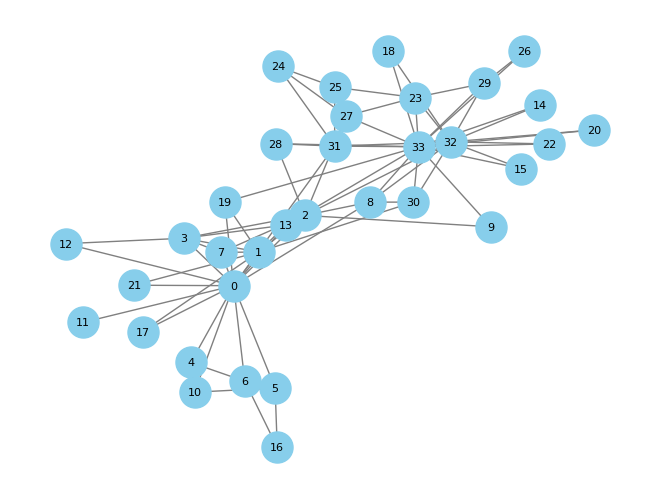}
    \caption{Spring Layout of a Sample Graph}
\end{figure}

This algorithm repetitively adjusts node positions based on repulsive and attractive forces, reaching an equilibrium where edge lengths are minimized and node overlap is reduced. The Fruchterman-Reingold algorithm, discussed by \cite{radovs2019fruchterman}, is widely applied in mathematics, computer science, and network visualization tasks for visually balanced graphs.

Each layout has unique strengths suited to different applications:
\begin{itemize}
    \item \textbf{Circular Layouts} highlight cyclic structures and symmetry, useful in hierarchical or modular systems.
    \item \textbf{Spectral Layouts} reveal clusters and assist in network segmentation, valuable for community or cluster visualization.
    \item \textbf{Spring Layouts} provide a balance between readability and spatial organization, ideal for undirected, unstructured graphs.
\end{itemize}

\subsection{Enhancing Graph Aesthetics}

\subsubsection{Using Colors to Enhance Readability}
Color is a fundamental tool in graph design, impacting both aesthetics and readability. Colors can highlight key areas, distinguish groups, and emphasize patterns without overwhelming the viewer.

\paragraph{Color Selection:} Choosing a cohesive color palette with appropriate contrast is essential for readability. For instance, contrasting colors (like dark blue on light yellow) improve visibility, while softer palettes work well for background elements to avoid distraction.

\paragraph{Applications in NetworkX:}
NetworkX allows for color customization using the \texttt{node\_color} and \texttt{edge\_color} parameters:
\begin{lstlisting}
import networkx as nx
import matplotlib.pyplot as plt

G = nx.cycle_graph(10)
colors = ["skyblue" if i % 2 == 0 else "salmon" for i in range(10)]
nx.draw(G, node_color=colors, with_labels=True)
plt.show()
\end{lstlisting}

This example alternates colors for nodes, which can help differentiate categories or clusters within a graph. According to \cite{butcherinfluence}, color choices impact not only aesthetic appeal but also viewer comprehension. Colors should be chosen thoughtfully to support the message and enhance user engagement without causing visual fatigue.

\subsubsection{Adding Labels for Clarity}
Labels provide essential context for nodes and edges, making it easier for viewers to understand the information conveyed in the graph. Labels should be legible, concise, and strategically positioned to avoid clutter.

\subsection{Labeling Nodes and Edges in NetworkX}
NetworkX supports adding labels to both nodes and edges, as shown below:

\begin{itemize}
    \item \textbf{Labeling Nodes and Edges in NetworkX}: NetworkX supports adding labels to both nodes and edges, as shown below:
\end{itemize}

\begin{lstlisting}[language=Python, caption={Labeling nodes and edges in a NetworkX graph}]
pos = nx.spring_layout(G)
nx.draw(G, pos, with_labels=True)
labels = {edge: f"{edge[0]}-{edge[1]}" for edge in G.edges()}
nx.draw_networkx_edge_labels(G, pos, edge_labels=labels)
plt.show()
\end{lstlisting}

\begin{figure}[h!]
    \centering
    \includegraphics[width=0.5\textwidth]{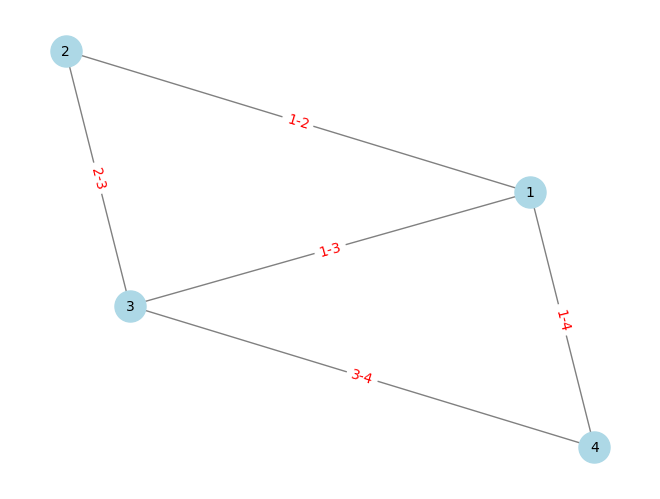}
    \caption{Graph with Labeled Nodes and Edges}
\end{figure}

Here, node labels are automatically placed based on node positions, and custom edge labels are defined for clarity. Studies on text aesthetics highlight that font choices and label placements significantly impact readability. A study by \cite{maity2018relating} demonstrated that font characteristics influence comprehension, which applies to labels in graphical interfaces as well. Using readable fonts, appropriate sizes, and avoiding overlap with nodes or edges enhances the viewer’s experience.

\subsubsection{Adding Legends for Context}
Legends are crucial in multi-colored or multi-symbol graphs, providing a reference for interpreting different colors, shapes, or line types. They allow viewers to quickly identify categories, improving the overall usability of the visualization.

\paragraph{Adding Legends in NetworkX:}
While NetworkX does not natively support legends, they can be added using \texttt{matplotlib} for clarity in complex graphs.

\begin{lstlisting}[language=Python, caption={Adding custom legends to a NetworkX graph}]
from matplotlib.lines import Line2D

# Define custom legends
legend_elements = [
    Line2D([0], [0], marker='o', color='w', label='Category 1', markerfacecolor='skyblue', markersize=10),
    Line2D([0], [0], marker='o', color='w', label='Category 2', markerfacecolor='salmon', markersize=10)
]
plt.legend(handles=legend_elements, loc="upper right")
plt.show()
\end{lstlisting}

\begin{figure}[h!]
    \centering
    \includegraphics[width=0.5\textwidth]{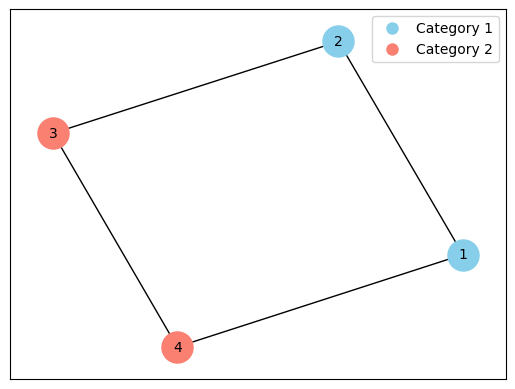}
    \caption{Graph with Custom Legend}
\end{figure}

This custom legend explains node colors, guiding viewers in interpreting categories or clusters. Proper legend placement and clarity are essential for readability. \cite{lulseged2018emj} emphasizes that a well-positioned, concise legend helps users interpret complex data representations without confusion. Legends should avoid excessive detail and be strategically placed, typically outside the main graph area, to prevent obstruction of key visuals.

\subsubsection{Combining Aesthetics for Optimal Graph Design}
For a graph to be effective, colors, labels, and legends must work together seamlessly. This enhances understanding by highlighting key areas, structuring information, and making data relationships intuitive, ultimately reducing cognitive load. \cite{lin2021aesthetics} discusses the importance of minimizing visual clutter by avoiding excessive edge crossings and ensuring that design choices, such as color schemes and grid patterns, align with user-friendly aesthetics for structured and accessible visual layouts.

\subsection{Interactive Graph Visualizations with Neo4j Bloom and Plotly}

\subsubsection{Interactive Visualization in Neo4j Bloom}
Neo4j Bloom is a visualization tool specifically designed for exploring data stored in Neo4j databases. It provides an interactive environment where users can investigate nodes and relationships in a visually rich manner.

\paragraph{Hover Effects} 
Hover effects in Neo4j Bloom display details about nodes and relationships when the cursor hovers over them. This feature is particularly useful for data exploration, as users can quickly view metadata or attributes without needing to click on each item. Simply move the cursor over any element in the Bloom visualization, and a popup will display relevant data, such as node type, properties, and relationship details.

\paragraph{Zoom and Pan} 
Neo4j Bloom includes zoom and pan features, enabling users to navigate large graphs effortlessly. With the mouse wheel or pinch gestures, users can zoom in on specific areas to inspect dense clusters or zoom out to get an overview of the entire network. Pan functionality allows users to click and drag across the graph, focusing on different sections without losing context.

\paragraph{Cypher Commands for Bloom Visualizations} 
Neo4j Bloom allows users to specify queries in Cypher to control the visualization, focusing on specific subsets of the graph:
\begin{lstlisting}
MATCH (p:Person)-[:FRIENDS_WITH]->(f:Person)
RETURN p, f
\end{lstlisting}
This query fetches only the nodes and relationships relevant to a specific pattern, streamlining the visualization and making it more manageable for exploration.

\paragraph{Benefits} 
Interactive features in Neo4j Bloom improve user engagement by making it easier to focus on specific graph segments. Studies highlight the effectiveness of interactive visualizations in fostering data exploration, particularly in network and relationship-rich datasets, where interactive tools allow deeper insights into hidden connections \cite{lulseged2018emj}.

\subsubsection{Interactive Visualization in Plotly}
Plotly is a Python-based visualization library that supports interactive features, including hover effects, zoom, and pan, making it ideal for exploratory data analysis of network graphs.

\paragraph{Creating a Graph with Plotly} 
Using Plotly’s scatter or NetworkX integration, you can create a network graph and enable interactive capabilities:

\begin{lstlisting}[language=Python, caption={Creating an interactive network graph with Plotly}]
import plotly.graph_objects as go
import networkx as nx

# Create a sample NetworkX graph
G = nx.cycle_graph(10)
pos = nx.spring_layout(G)

# Extract positions and create scatter plot traces
edge_x = []
edge_y = []
for edge in G.edges():
    x0, y0 = pos[edge[0]]
    x1, y1 = pos[edge[1]]
    edge_x.extend([x0, x1, None])
    edge_y.extend([y0, y1, None])
edge_trace = go.Scatter(x=edge_x, y=edge_y, line=dict(width=0.5, color="#888"),
                        hoverinfo="none", mode="lines")

node_x = [pos[node][0] for node in G.nodes()]
node_y = [pos[node][1] for node in G.nodes()]
node_trace = go.Scatter(x=node_x, y=node_y, mode="markers+text",
                        hoverinfo="text", marker=dict(size=10, color="skyblue"))

fig = go.Figure(data=[edge_trace, node_trace])
fig.update_layout(showlegend=False)
fig.show()
\end{lstlisting}

This code example builds an interactive network plot, where nodes and edges are visually represented, and hover information can be customized to show additional metadata about each node.

\paragraph{Hover Effects} 
Hover effects in Plotly allow users to quickly access information about nodes, such as their labels or connected nodes. Custom hover text can be added to each node in \texttt{node\_trace} using the \texttt{text} parameter, which can include node attributes or other relevant details:

\begin{lstlisting}
node_trace.text = [f"Node {node}" for node in G.nodes()]
\end{lstlisting}

\paragraph{Zoom and Pan} 
Plotly’s zoom and pan features are automatically enabled in interactive graphs. Users can zoom by scrolling or pinching and pan by clicking and dragging. This functionality is valuable for detailed analysis, allowing users to seamlessly navigate large networks.

\subsubsection{Benefits of Interactive Visualization in Graph Analysis}
Interactive visualization enhances data analysis by allowing users to engage dynamically with the data, fostering insights that static visualizations may obscure. Interactive features like hover effects, zoom, and pan are essential for exploring complex networks, as they reveal details only when necessary, preserving visual clarity.

Studies confirm that interactive visualizations significantly improve the user experience, making it easier to detect patterns and gain insights. \cite{lin2021aesthetics} found that interactivity, such as zoom and hover details, enhances the clarity of network visualizations by enabling users to focus on relevant sections without information overload.

\section{Advanced Graph Algorithms}
\subsection{Community Detection and Clustering}
\subsubsection{Louvain Algorithm}
The Louvain algorithm is a popular method for community detection in large networks due to its scalability and effectiveness in maximizing modularity. This algorithm operates in two phases: modularity optimization, where nodes are grouped into communities, and community aggregation, where each community is treated as a single node. These phases repeat iteratively until the modularity reaches its peak.

\paragraph{Implementation in Python with NetworkX}
NetworkX offers an implementation of the Louvain algorithm through external libraries like \texttt{community}:

\begin{lstlisting}[language=Python, caption={Louvain community detection in a NetworkX graph}]
import networkx as nx
import community as community_louvain
import matplotlib.pyplot as plt
G = nx.erdos_renyi_graph(100, 0.05)
# Compute the best partition using Louvain
partition = community_louvain.best_partition(G)
# Color nodes based on community
pos = nx.spring_layout(G)
cmap = plt.cm.get_cmap("viridis", max(partition.values()) + 1)
nx.draw_networkx_nodes(G, pos, partition.keys(), node_size=40, 
                       cmap=cmap, node_color=list(partition.values()))
nx.draw_networkx_edges(G, pos, alpha=0.5)
plt.show()
\end{lstlisting}

\begin{figure}[h!]
    \centering
    \includegraphics[width=0.5\textwidth]{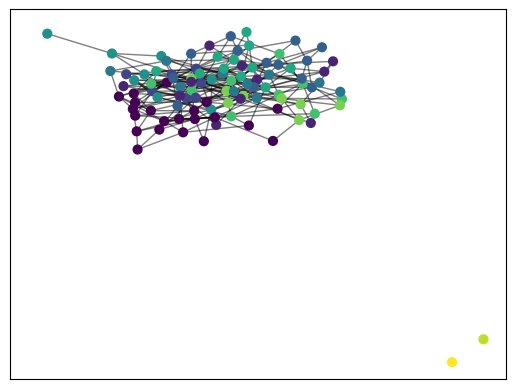}
    \caption{Louvain Community Detection in a NetworkX Graph}
\end{figure}

This example uses the Louvain method to partition a random graph, coloring nodes according to detected communities. The Louvain algorithm has broad applications in social network analysis, where identifying community structures helps understand user groupings or detect hidden relationships. Recent work, such as by \cite{inuwa2021multilevel}, applies community detection in social media to enhance recommendation systems by grouping users with similar behavior patterns.

\subsubsection{Other Community Detection Algorithms}
Several other community detection algorithms complement the Louvain method, each suited to specific types of networks or goals:

\begin{itemize}
    \item \textbf{Degenerate Agglomerative Hierarchical Clustering Algorithm (DAHCA)}: DAHCA clusters nodes based on vertex similarity, often outperforming traditional methods in networks with low intra-community connectivity, particularly in biological and social networks \cite{maria2020vertex}.
    \item \textbf{Hierarchical Clustering}: Effective for analyzing complex, multi-level community structures, particularly in large datasets where computational complexity is a consideration \cite{thai2023intelligent}.
    \item \textbf{Overlapping Community Detection}: Allows nodes to belong to multiple communities, reflecting real-world networks where entities often belong to different social or functional groups \cite{grass2020overlapping}.
\end{itemize}

\subsubsection{Applications of Community Detection in Real-World Networks}
Community detection algorithms are widely applied across various fields:

\begin{itemize}
    \item \textbf{Social Networks}: Identifies influential groups, understands social dynamics, and improves targeted recommendations in social media platforms \cite{deng2017efficient}.
    \item \textbf{Biological Networks}: In genomics, detecting clusters of interacting genes or proteins can reveal insights into biological functions and disease mechanisms.
    \item \textbf{Collaborative Filtering}: Enhances recommendation systems by identifying clusters of users with similar preferences.
\end{itemize}

\subsection{Centrality Measures in Network Analysis}

\subsubsection{Degree Centrality}
Degree centrality is defined as the number of direct connections a node has, reflecting a node's immediate influence within the network.

\paragraph{Formula:} 
\[ C_D(v) = \text{deg}(v) \]

\paragraph{Implementation in NetworkX} \hfill \\ 

\paragraph{Calculating Degree Centrality in NetworkX}
Degree centrality measures the fraction of nodes a particular node is connected to, highlighting its local importance within the network.

\begin{lstlisting}[language=Python, caption={Calculating degree centrality of nodes in a NetworkX graph}]
import networkx as nx
G = nx.Graph()
G.add_edges_from([(1, 2), (2, 3), (3, 4), (4, 5), (1, 5)])
degree_centrality = nx.degree_centrality(G)
print(degree_centrality)
\end{lstlisting}

\begin{tcolorbox}[colback=gray!10, colframe=gray!80!black, title=Expected Output]
{
1: 0.5, \\
2: 0.5, \\
3: 0.5, \\
4: 0.5, \\
5: 0.5
}
\end{tcolorbox}

\subsubsection{Closeness Centrality}
Closeness centrality measures how close a node is to all other nodes in the network.

\paragraph{Formula:} 
\[
C_C(v) = \frac{1}{\sum_{u \neq v} d(u, v)}
\]
where \( d(u, v) \) is the shortest path distance between \( u \) and \( v \).

\paragraph{Implementation in NetworkX}\hfill \\ 
\begin{lstlisting}[language=Python, caption={Calculating closeness centrality in a NetworkX graph}]
closeness_centrality = nx.closeness_centrality(G)
print(closeness_centrality)
\end{lstlisting}

\begin{tcolorbox}[colback=gray!10, colframe=gray!80!black, title=Expected Output]
{
1: 0.67, \\
2: 0.80, \\
3: 1.00, \\
4: 0.80, \\
5: 0.67
}
\end{tcolorbox}

\subsubsection{Betweenness Centrality}
Betweenness centrality quantifies the extent to which a node lies on the shortest paths between other nodes.

\paragraph{Formula:} 
\[
C_B(v) = \sum_{s \neq v \neq t} \frac{\sigma_{st}(v)}{\sigma_{st}}
\]
where \( \sigma_{st} \) is the total number of shortest paths from node \( s \) to \( t \), and \( \sigma_{st}(v) \) is the number of those paths that pass through \( v \).

\paragraph{Implementation in NetworkX}\hfill \\ 
\begin{lstlisting}
betweenness_centrality = nx.betweenness_centrality(G)
print(betweenness_centrality)
\end{lstlisting}

\subsubsection{Eigenvector Centrality}
Eigenvector centrality measures a node’s influence based on its connections, considering the importance of its neighbors.

\paragraph{Formula:} 
\[
C_E(v) = \frac{1}{\lambda} \sum_{u \in N(v)} C_E(u)
\]
where \( N(v) \) are the neighbors of \( v \) and \( \lambda \) is the largest eigenvalue of the adjacency matrix.

\paragraph{Implementation in NetworkX} \hfill \\ 
\begin{lstlisting}
eigenvector_centrality = nx.eigenvector_centrality(G)
print(eigenvector_centrality)
\end{lstlisting}

\subsection{PageRank and Similar Ranking Algorithms}

\subsubsection{PageRank Algorithm}
The PageRank algorithm assigns a ranking to each node based on the number and quality of links to it.

\paragraph{Mathematical Formula:} 
\[
PR(v) = \frac{1 - d}{N} + d \sum_{u \in M(v)} \frac{PR(u)}{L(u)}
\]
where \( d \) is the damping factor, \( N \) is the total number of nodes, \( M(v) \) is the set of nodes that link to \( v \), and \( L(u) \) is the number of outbound links from \( u \).

\paragraph{Implementation in NetworkX} \hfill \\
The PageRank algorithm evaluates the importance of each node based on the structure of inbound links, commonly used in search engines to rank web pages.

\begin{lstlisting}[language=Python, caption={Calculating PageRank in a NetworkX graph}]
import networkx as nx
G = nx.DiGraph()
G.add_edges_from([(1, 2), (2, 3), (3, 1), (3, 4), (4, 2)])
pagerank_scores = nx.pagerank(G, alpha=0.85)
print("PageRank Scores:", pagerank_scores)
\end{lstlisting}

\begin{tcolorbox}[colback=gray!10, colframe=gray!80!black, title=Expected Output]
PageRank Scores: \\
{
1: 0.29, \\
2: 0.34, \\
3: 0.26, \\
4: 0.11
}
\end{tcolorbox}

\subsubsection{Weighted PageRank} \hfill \\ 
The Weighted PageRank (WPR) algorithm incorporates the importance of inbound and outbound links, adjusting scores based on the weight of each edge.

\paragraph{Implementation in NetworkX} \hfill \\
\begin{lstlisting}[language=Python, caption={Calculating Weighted PageRank in a NetworkX graph}]
G.add_weighted_edges_from([(1, 2, 0.8), 
(2, 3, 0.5), (3, 1, 0.6)])
weighted_pagerank_scores = nx.pagerank(G, alpha=0.85, weight='weight')
print("Weighted PageRank Scores:", weighted_pagerank_scores)
\end{lstlisting}

\begin{tcolorbox}[colback=gray!10, colframe=gray!80!black, title=Expected Output]
Weighted PageRank Scores: \\
{
1: 0.30, \\
2: 0.33, \\
3: 0.25
}
\end{tcolorbox}

\subsubsection{Topic-Sensitive PageRank} \hfill \\ 
Topic-Sensitive PageRank (TSPR) ranks nodes based on relevance to specific topics, making it suitable for personalized content delivery.

\subsubsection{Hybrid and Advanced Ranking Methods} \hfill \\ 
Recent advancements integrate structural and user behavior metrics in ranking methods, improving relevance for recommendation systems \cite{agnihotri2024iot, zhuimproved}.

\section{Applications of Graph Databases in Data Science}

\subsection{Social Network Analysis (SNA)}

Social Network Analysis (SNA) is an essential approach in data science for understanding the structure and dynamics of social interactions. By analyzing the patterns and relationships within social networks, SNA can help identify key influencers and discover community clusters. Neo4j, a graph database platform, provides powerful tools for conducting SNA, including influencer detection and community clustering.

\subsubsection{Basics of Social Network Analysis}
SNA examines the nodes (individuals or entities) and edges (relationships) within a network to identify patterns in connectivity and influence. Core metrics used in SNA include:
\begin{itemize}
    \item \textbf{Degree Centrality}: Measures the number of direct connections a node has, helping identify well-connected individuals within a network.
    \item \textbf{Betweenness Centrality}: Quantifies how often a node acts as a bridge along the shortest path between other nodes, marking those who facilitate communication between clusters.
    \item \textbf{PageRank}: An algorithm developed by Google to rank web pages, adapted in SNA to evaluate influence based on the importance of a node’s connections \cite{farooq2018detection}.
\end{itemize}

\subsubsection{Identifying Influencers with Neo4j}
In Neo4j, influencers within a network can be identified by calculating centrality metrics.

\paragraph{Cypher Query for Degree Centrality} \hfill \\ 
\begin{lstlisting}
MATCH (p:Person)-[:FRIENDS_WITH]-(f)
RETURN p.name AS Name, COUNT(f) AS DegreeCentrality
ORDER BY DegreeCentrality DESC
\end{lstlisting}

This query lists individuals with the most connections, highlighting potential influencers based on their direct links \cite{mittal2019classifying}. 
\\

\subsubsection{PageRank in Neo4j}
This PageRank query calculates the influence of nodes within a graph, ranking users based on the importance of their connections. \\

\begin{lstlisting}
CALL gds.pageRank.stream('myGraph')
YIELD nodeId, score
RETURN gds.util.asNode(nodeId).name AS Name, score
ORDER BY score DESC
\end{lstlisting}

\subsubsection{Community Detection for Clustering}
Community detection algorithms in Neo4j can identify clusters or groups within a network, often revealing hidden structures.

\paragraph{Louvain Algorithm for Community Detection}
Neo4j’s implementation of the Louvain algorithm helps in detecting clusters by optimizing modularity. This approach is effective in social networks for identifying communities with shared interests or similar attributes \cite{zhou2019research}.
\begin{lstlisting}
CALL gds.louvain.stream('myGraph')
YIELD nodeId, communityId
RETURN gds.util.asNode(nodeId).name AS Name, communityId
ORDER BY communityId
\end{lstlisting}

\subsubsection{Applications of SNA}
Social network analysis can greatly benefit fields such as marketing, epidemiology, and organizational management by providing insights into network structures and behaviors. For example, influencer identification supports marketing efforts by targeting users with high impact, while cluster detection facilitates community building and content recommendation \cite{zhou2019research}.

\subsection{Recommender Systems with Graphs}

\subsubsection{Graph-Based Collaborative Filtering (CF)}
Graph-based collaborative filtering (CF) leverages user-item interactions in graph structures to enhance recommender systems. By modeling users and items as nodes and their relationships (e.g., ratings or interactions) as edges, graph-based CF effectively captures complex, multi-level connections in data, addressing data sparsity and cold-start issues found in traditional CF methods.

\paragraph{Graph-Based Collaborative Filtering Methods}
\begin{itemize}
    \item \textbf{Basic Graph-Based CF}: Represents user-item interactions in a bipartite graph and applies neighborhood-based methods to identify similarities \cite{satheesh2017collaborative}.
    \item \textbf{Graph Signal Processing for CF}: Integrates graph signal processing to refine CF models, treating user-item graphs as signal domains \cite{huang2017collaborative}.
    \item \textbf{Graph-Aware Collaborative Filtering}: Incorporates both user-item bipartite graphs and knowledge graphs to refine representations \cite{long2021graph}.
    \item \textbf{Graph Neural Networks (GNN) for CF}: Uses GNNs to apply neighborhood aggregation, allowing the model to learn representations from multi-hop neighbors \cite{gwadabe2022improving, gilmer2017neural}.
\end{itemize}

\paragraph{Implementation with Neo4j} 
Neo4j can be used to build graph-based recommendation engines by executing queries to identify similar items or recommend items based on user connections:

\begin{lstlisting}[language=, caption={Building a recommendation engine in Neo4j}]
    MATCH (u:User)-[r:RATED]->(m:Movie)
    WITH u, collect(m) AS movies
    MATCH (u2:User)-[r:RATED]->(m2:Movie)
    WHERE u <> u2 AND m2 IN movies
    RETURN u2.name AS SimilarUser, count(*) AS SharedInterests
    ORDER BY SharedInterests DESC
    LIMIT 5
\end{lstlisting}

\begin{tcolorbox}[colback=gray!10, colframe=gray!80!black, title=Expected Output]
\begin{tabular}{|c|c|}
\hline
\textbf{SimilarUser} & \textbf{SharedInterests} \\
\hline
Alice & 5 \\
Bob   & 4 \\
Carol & 3 \\
David & 3 \\
Eve   & 2 \\
\hline
\end{tabular}
\end{tcolorbox}

This Cypher query identifies users with shared interests by counting common items they’ve rated, aiding in recommending items based on similar user profiles.

\subsubsection{Applications of Graph-Based CF}
Graph-based CF is widely used in e-commerce, content streaming, and social media due to its flexibility and accuracy in handling diverse data types. This approach allows CF systems to incorporate heterogeneous data, such as social connections or contextual knowledge, which improves recommendation quality and addresses data sparsity \cite{mu2023heterogeneous}.

\subsection{Anomaly Detection Using Graphs}

Graph-based anomaly detection is crucial for identifying irregular patterns in network structures, particularly in fraud detection, where suspicious transactions deviate from normal patterns.

\subsubsection{Graph-Based Anomaly Detection Techniques}
Anomaly detection in graphs involves analyzing network structures to identify outlier nodes or edges.

\begin{itemize}
    \item \textbf{Structural Anomaly Detection}: Identifies nodes or edges with unusual degrees or connection patterns, often signaling fraudulent behavior \cite{nguyen2023example}.
    \item \textbf{Community-Based Detection}: Detects anomalies within communities, identifying outliers in otherwise cohesive groups \cite{bukhori2023inductive}.
\end{itemize}

\subsubsection{Graph Neural Networks (GNNs)}
GNNs are powerful tools for anomaly detection, capturing both structure and attributes of nodes. For example, \cite{motie2024financial} reviewed GNN applications in financial fraud detection, where GNNs identify complex anomalies across transaction networks by aggregating neighborhood information.

\subsubsection{Case Study: Fraud Detection with Graph Analysis}
In financial networks, fraud detection can be achieved by representing entities as nodes and transactions as edges.

\paragraph{Data Fusion and Graph Analysis}
A study by \cite{bellandi2022data} used data fusion and graph analysis to detect fraudulent transactions by integrating transaction metadata with connection patterns. \hfill \\ 

\paragraph{Cypher Query in Neo4j for Fraud Detection} \hfill \\
\begin{lstlisting}
MATCH (a:Account)-[t:TRANSFER]->(b:Account)
WHERE t.amount > 10000 AND a.region <> b.region
RETURN a, b, t.amount
\end{lstlisting}
This query identifies large transactions between accounts in different regions, often associated with money laundering or fraud.

\paragraph{Graph Neural Networks with Self-Attention}
\cite{li2021fraud} applied a GNN with self-attention to detect fraud in electronic payment systems, combining transaction characteristics with network structure for improved detection.

\subsubsection{Benefits of Graph-Based Anomaly Detection}
Graph-based anomaly detection offers distinct advantages:
\begin{itemize}
    \item \textbf{Handling Complex Relationships}: Graphs naturally represent relationships, making fraud detection more accurate in highly interconnected networks.
    \item \textbf{Adaptability to Dynamic Data}: Graph databases like Neo4j support real-time anomaly detection, adapting as new data arrives.
    \item \textbf{Enhanced Accuracy with Machine Learning}: GNNs and other advanced models improve detection precision by leveraging complex patterns in structure and attributes \cite{tran2022some}.
\end{itemize}

\section{Working with Large Graphs and Optimization Techniques}

\subsection{Handling Large Graphs in Python and Neo4j}

For handling large graphs in Python and Neo4j, memory optimization is essential to manage data efficiently, especially given the complex and often irregular structures of large graphs.

\subsubsection{Memory Optimization Techniques}
\begin{itemize}
    \item \textbf{Compressed Data Structures}: Techniques like compressed adjacency lists and variable-byte encoding can reduce memory usage by up to 5x while preserving access efficiency. By compressing edge weights and node attributes, these methods minimize memory demands, particularly useful in Python-based systems like NetworkX \cite{liakos2017realizing}.
    \item \textbf{Graph Partitioning}: This method breaks large graphs into smaller subgraphs, each fitting into memory individually. The GraphH system, for example, uses an edge cache mechanism to reduce disk I/O overhead, optimizing memory for large graphs, particularly when partitioning nodes and edges in Neo4j \cite{sun2017graphh}.
    \item \textbf{Vertex Merging and Compression}: Techniques like Vertex Merging (VM) group similar vertices to reduce redundant data processing. Aggressive vertex merging further optimizes memory by minimizing dependencies between data points, speeding up execution time \cite{cheng2022graphrc}.
\end{itemize}

\subsubsection{Advanced Graph Libraries and Frameworks}
\begin{itemize}
    \item \textbf{Graph-XLL}: This library is optimized for large graphs on consumer-grade machines, significantly reducing memory requirements. It offers scalable graph analytics that helps prevent out-of-memory errors common with NetworkX in Python \cite{wu2019graph}.
    \item \textbf{Neo4j Memory Management}: Neo4j uses memory mapping and efficient page caching to handle large graphs. Its graph engine manages data that exceeds memory capacity through optimized I/O management, supporting bulk ingestion with near-storage accelerators \cite{kang2023near}.
\end{itemize}

\subsubsection{Out-of-Core Processing and Specialized Hardware}
For graphs that exceed in-memory capabilities, out-of-core processing enables systems to offload data to disk, processing only necessary portions in memory. GraphMP, for instance, uses a vertex-centric sliding window model and selective scheduling to minimize memory consumption \cite{sun2017graphh}.

Specialized hardware, like FPGA boards, can handle memory bottlenecks through parallel processing, embedding graph structures into silicon to reduce memory access latency \cite{mokhov2017language}. These techniques support large-scale applications like social network analysis, recommendation systems, and fraud detection.

\subsection{Graph Databases for Large-Scale Applications}

Graph databases like Neo4j are essential for large-scale applications, particularly when managing complex, interconnected datasets where traditional databases struggle. Neo4j supports efficient data loading, querying, and management of graph data.

\subsubsection{Data Loading in Neo4j}
Efficient data loading is crucial in Neo4j, especially for large applications involving high volumes of interconnected data. Neo4j utilizes batch data ingestion and parallel processing to enhance data import speed. For example, large-scale biological networks benefit from Neo4j's efficient data loading capabilities \cite{fabregat2018reactome}.

\subsubsection{Query Optimization and Execution}
Neo4j provides optimized query execution for handling complex graph patterns. Its Cypher query language enables efficient traversal and pattern matching, and its caching mechanism reduces database access during high-frequency queries. A comparative study by \cite{do2022query} found that Neo4j outperformed relational databases like MySQL in query execution time, demonstrating its effectiveness for social network analysis and recommendation systems.

\paragraph{Cypher Query Example:}\hfill \\ 
\begin{lstlisting}
MATCH (p:Person)-[:FRIEND]->(f:Person)
WHERE p.age > 30
RETURN f.name
\end{lstlisting}
This query finds friends of people over 30, showcasing Neo4j’s efficiency in traversing relationships quickly.

\subsubsection{Data Management and Scaling}
Neo4j leverages physical RAM to enhance data access speeds and integrates disk-based storage when datasets exceed memory capacity. A study comparing Neo4j and Apache Spark highlighted Neo4j’s capacity to manage large data until reaching memory limits, at which point Spark may take over for distributed processing needs \cite{ballas2020assessing}.

\subsubsection{Applications and Benefits in Large-Scale Environments}
Graph databases excel in fields such as social networks, healthcare, and bioinformatics, where data relationships are complex and flexible querying is required. For instance, Neo4j is used in bioinformatics to manage biomolecular pathway data, achieving up to 93\% faster query performance than relational databases \cite{fabregat2018reactome}. Neo4j’s support for Enhanced Entity-Relationship (EER) models further validates its scalability \cite{vagner2018store}.

\subsection{Parallel Processing and Performance Optimization}

Parallel processing in Neo4j, combined with code optimization techniques, significantly enhances performance for large-scale applications requiring efficient resource utilization and rapid query execution.

\subsubsection{Parallel Processing in Neo4j}
Neo4j supports parallel processing to improve data loading, querying, and transaction handling. Parallel Cypher Execution allows Neo4j to split queries into sub-tasks processed concurrently, increasing throughput and reducing query times. Neo4j also employs multi-threading for batch imports and complex traversals, supporting applications like fraud detection and recommendation systems \cite{liu2018performance}.

\paragraph{Batch Loading with Parallel Processing:}\hfill \\ 
\begin{lstlisting}
CALL apoc.periodic.iterate(
  "LOAD CSV WITH HEADERS FROM 'file:///large_data.csv' AS row RETURN row",
  "CREATE (:Entity {id: row.id, name: row.name, value: row.value})",
  {batchSize: 1000, parallel: true}
)
\end{lstlisting}
This command loads data in parallel with a batch size of 1000, reducing memory overhead and optimizing CPU utilization.

\subsubsection{Code Optimization Techniques}
Code optimization in Neo4j involves tuning Cypher queries to reduce computational load and memory use. Techniques such as query restructuring and index usage can drastically reduce execution times. For instance, creating indexes on frequently queried properties enhances retrieval speed.

\paragraph{Optimization Example with Indexes:}\hfill \\ 
\begin{lstlisting}
CREATE INDEX FOR (n:Entity) ON (n.name)
\end{lstlisting}
This command sets up an index on the name property of Entity nodes, accelerating queries that filter by name.

\subsubsection{Advanced Parallel Processing Techniques and Hardware Utilization}
For further efficiency, Neo4j can integrate with specialized parallel processing architectures, such as multi-GPU environments or FPGA boards. Studies show that combining GPUs and FPGAs for graph data processing can yield significant performance improvements \cite{wang2019optimization}.

\paragraph{Parallelism in Graph Traversals:}
Optimization algorithms employing loop unrolling and dynamic scheduling enhance Neo4j’s graph traversal performance. Loop unrolling, in particular, is effective for tasks like Sparse Matrix Multiplications, applicable to graph traversal optimization \cite{soliman2018loop}.

\subsubsection{Performance Studies and Results}
Recent studies have demonstrated substantial performance improvements through parallelism and code optimization in Neo4j. For example, a hybrid parallel approach using OpenMP and MPI with Neo4j has shown to accelerate processing times by up to 3.2 times compared to conventional methods \cite{borovska2017hybrid}.

\section{Building a Real-World Graph-Based Project}

\subsection{Planning a Graph-Based Project}

To plan a graph-based project effectively, attention must be given to objective setting, project structure, and algorithm selection, which are crucial for optimizing project outcomes and resource utilization.

\subsubsection{Objective Setting}
Objective setting is foundational in graph-based projects, as it defines the structure and requirements for the graph model. Objectives may include optimizing data processing speed, minimizing project costs, or maximizing algorithm accuracy. Pareto optimization can be used to balance competing objectives, such as efficiency and accuracy, by identifying solutions that achieve the best trade-offs \cite{bugaeev2017algorithm}. Additionally, graph-analytic models can aid in structuring project objectives and evaluating planning methods \cite{tseluyko2023study}.

\subsubsection{Algorithm Selection}
Choosing the right algorithms is critical for achieving project objectives. The selection process should consider factors such as data structure, computational complexity, and specific project goals. A machine learning-based algorithm selection approach can dynamically choose optimal graph partitioning strategies, enhancing performance on large datasets \cite{park2022machine}. For instance, the Pregel algorithm is effective for scalable computations in distributed environments, making it suitable for projects that require extensive parallel processing \cite{jiang2019scalable}.

\subsubsection{Graph Models and Optimization Techniques}
Graph models in engineering and economic planning represent complex relationships, with graph rewriting techniques optimizing design processes. This method enables formalized rule-based transformations, supporting objective setting and algorithm selection \cite{kolbeck2022graph}. Additionally, data, information, and knowledge graphs enhance technical and economic planning by aligning software design objectives with implementation \cite{shao2017bidirectional}.

\subsubsection{Applied Algorithm Strategies}
For large-scale applications, the LAGraph project assembles verified algorithms into a unified framework, aiding in efficient and predictable graph processing, especially valuable during the planning phase \cite{mattson2019lagraph}. Predictive workload models, such as Graph-Optimizer, help estimate performance needs to ensure selected algorithms align with project objectives \cite{varbanescu2023graph}.

\subsection{Implementing a Graph Database Solution (Recommender System or Social Network Analysis Tool)}

To implement a graph database solution like a recommender system in Neo4j, particularly one integrated with machine learning, follow these steps:

\subsubsection{Data Collection and Preprocessing}
Data preparation is essential for effective graph representation. Collect data relevant to the recommendation criteria, such as user preferences or historical interactions. Store this data as nodes (e.g., users, items) and relationships (e.g., purchases, likes) in Neo4j. Data can be ingested using Python or ETL processes and structured for graph processing \cite{bajaj2019species}.

\subsubsection{Data Modeling in Neo4j}
Define the schema in Neo4j, optimizing it for querying. For a recommender system, nodes represent entities like users and products, while relationships capture interactions. Neo4j's Cypher query language enables flexible schema design and data modification:
\begin{lstlisting}
CREATE (u:User {id: 1, name: "Alice"})
CREATE (p:Product {id: 101, name: "Product A"})
CREATE (u)-[:PURCHASED]->(p)
\end{lstlisting}
This basic structure facilitates recommendations based on relationship patterns \cite{henna2021enterprise}.

\subsubsection{Integrating Machine Learning Models}
Integrate machine learning models that leverage the graph data. Algorithms like collaborative filtering or deep learning methods such as graph neural networks (GNNs) can enhance recommendations by analyzing patterns in user interactions. Data-centric Graph ML techniques support improved graph representation and pattern recognition \cite{zheng2023towards}.

\subsubsection{Implementing Recommendation Algorithms in Neo4j}
Neo4j supports in-database machine learning through plugins and procedures, such as implementing a Decision Tree model within the database. With the \texttt{gds} (Graph Data Science) library functions, specific algorithms can be employed to achieve advanced recommendations.

\subsection{Deploying Graph Applications}

Deploying graph applications with user-friendly interfaces requires prioritizing usability, interactivity, and scalability, especially when handling large datasets and complex analytics.

\subsubsection{Designing Interactive User Interfaces}
User-friendly interfaces broaden the accessibility of graph applications, facilitating exploratory and analytical tasks. For example, \cite{erra2018methodological} showcases a 3D immersive environment for graph exploration, enhancing user engagement through virtual reality. These interfaces use intuitive controls like point-and-click or drag-and-drop, enabling non-technical users to navigate complex graph data effortlessly.

\subsubsection{Web-Based and Visual Tools for Deployment}
Web-based platforms offer accessibility and interactivity for graph applications. \cite{ivanoska2019web} developed a web tool that allows users to compare graph embedding techniques for tasks like node classification, supporting non-experts in selecting suitable graph representations. This platform demonstrates how web-based solutions can improve user experience by embedding graph algorithms and visualizations in a browser-friendly format.

\subsubsection{Edge-Cloud Collaboration for Scalable Solutions}
An edge-cloud framework optimizes graph processing for large-scale deployments, enhancing performance and responsiveness while allowing distributed data storage. \cite{zhou2023edge} presents an edge-cloud collaboration model, enabling graph applications to manage heavy computational loads at the cloud level while providing a responsive user experience on edge devices.

\subsubsection{Declarative Graph Analytics with Gradoop}
Gradoop, an open-source framework, enables users to define and execute distributed graph analysis programs using a high-level language. It supports pattern matching and graph grouping with a user-friendly interface, reducing the need for advanced coding skills, as highlighted by \cite{junghanns2018declarative}. This approach streamlines complex graph operations, facilitating efficient data handling and scalability across large datasets.

\subsubsection{Implementing Real-Time Visualization Frameworks}
Real-time visualization is critical for user engagement, especially for scenarios requiring dynamic data updates. \cite{komaromi2018efficient} implemented a framework in RefactorErl for visualizing large Semantic Program Graphs in near real-time, helping users understand structural changes as they occur. This approach can also be applied to real-time graph applications, such as social network monitoring.

\subsubsection{Deploying Customizable Graph Processing Tools}
GraphScope, a Python-based engine for distributed graph processing, offers a customizable environment for large-scale graph analytics \cite{fan2021graphscope}. It integrates with a Python API, allowing users to configure and run graph algorithms interactively, which enhances deployment flexibility for developers needing adaptable graph solutions.

\section{Advanced Topics and Further Reading}

\subsection{Graphs in Natural Language Processing (NLP)}

In Natural Language Processing (NLP), graphs play a significant role by organizing complex relationships between entities, making tasks like knowledge graph construction, text clustering, and relationship extraction more efficient and accurate.

\subsubsection{Knowledge Graphs in NLP}
Knowledge graphs are structured representations that capture entities as nodes and relationships as edges, structuring semantic relationships extracted from unstructured text. Constructing knowledge graphs enhances data extraction for applications like information retrieval and semantic search. For instance, Text2NKG introduces a fine-grained n-ary relation extraction framework, automating the capture of complex entity relations across various contexts \cite{luo2023text2nkg}. Additionally, knowledge graphs support feature extraction in text classification by integrating relational structures, as shown by the Microsoft Concept Graph \cite{petrvzelkova2020knowledge}.

\subsubsection{Graph-Based Text Clustering}
Graph-based text clustering uses connectivity between entities or keywords to group similar documents or text segments. Algorithms like CLIP use link features to enhance clustering quality in knowledge graphs, especially in scenarios with overlapping clusters \cite{saeedi2018using}. This approach benefits applications like social media analysis, where grouping text thematically provides valuable insights.

\subsubsection{Relationship Extraction with Graph Structures}
Relationship extraction is essential for building knowledge graphs and identifying connections between entities in texts. Graph-based frameworks often use Graph Neural Networks (GNNs) or Graph Attention Networks to contextualize relationships, enabling more accurate capture of relational data. For example, sentential relation extraction is enhanced by dual heterogeneous graph context selection, which leverages both graph structure and node features \cite{xu2023knowledge}.

\subsection{Graph Neural Networks (GNNs)}

Graph Neural Networks (GNNs) have become essential in network analysis, enabling the learning of representations from graph-structured data. GNNs iteratively pass information (messages) between nodes and edges, allowing each node to update its representation based on neighboring nodes. This process produces learned embeddings that capture local and global graph structure, supporting tasks such as node classification, link prediction, and graph classification.

\subsubsection{Basic Operation of GNNs}
The core function of GNNs is based on a message-passing mechanism:
\[
h^{(k)} = \text{Update}(h^{(k-1)}, \text{Aggregate}(\{h^{(k-1)}_u : u \in N(v)\}))
\]
where \( h^{(k)} \) is the hidden state (or embedding) of node \( v \) at layer \( k \), \( N(v) \) represents the neighbors of \( v \), and functions like \texttt{Aggregate} and \texttt{Update} define how information from neighboring nodes is combined. Graph Convolutional Networks (GCNs), a common GNN implementation, update node features by averaging neighbors' features weighted by edge values \cite{liu2023survey, mladenovic1997variable}.

\subsubsection{GNN Applications in Network Analysis}
\begin{itemize}
    \item \textbf{Node Classification}: GNNs are widely used in node classification tasks, where the goal is to predict labels for nodes based on graph structure and node attributes. Applications include classifying users in social networks or detecting malicious nodes in cybersecurity \cite{pfrommer2021discriminability}.
    \item \textbf{Link Prediction}: In link prediction, GNNs predict the likelihood of edges between nodes, enabling applications like friend recommendations or collaboration suggestions. DyGNN, a dynamic GNN model, captures temporal interactions for evolving graphs, making it suitable for predicting new connections over time \cite{ma2020streaming}.
    \item \textbf{Graph Classification}: Graph classification aims to categorize entire graphs, useful in domains like bioinformatics, where GNNs can classify molecular structures. Kernel-based GNNs (KGNNs) improve classification accuracy by incorporating both labeled and unlabeled data \cite{ju2022kgnn}.
\end{itemize}

\subsubsection{Advanced GNN Architectures and Optimization}
\begin{itemize}
    \item \textbf{Graph Convolutional Networks (GCNs)}: GCNs extend convolution operations to graph structures, aggregating information from neighboring nodes. This architecture is widely used in applications requiring relational information, such as traffic prediction and community detection.
    \item \textbf{Graph Attention Networks (GATs)}: GATs enhance GCNs by introducing attention mechanisms, allowing nodes to assign different weights to neighbors based on relevance. This capability is useful in environments like financial transaction monitoring where data noise is a concern \cite{duong2019node}.
    \item \textbf{Graph Neural Architecture Search (GraphNAS)}: GraphNAS automates GNN architecture design using reinforcement learning, optimizing network structures for tasks like node classification, reducing manual design efforts \cite{gao2019graphnas}.
\end{itemize}

\subsection{Future of Graph Theory in Data Science}

Graph theory is evolving rapidly within data science, providing innovative approaches for data representation, analysis, and prediction across various domains. Key trends include graph embedding, dynamic graph modeling, graph databases, and graph signal processing.

\subsubsection{Graph Embedding and Representation Learning}
Graph embedding transforms complex graph structures into vector representations, allowing them to be processed by machine learning algorithms. This technique is essential in applications like social network analysis, recommendation systems, and biological data modeling, as embeddings retain relational and structural graph information \cite{cai2018comprehensive}.

\subsubsection{Temporal and Dynamic Graph Modeling}
Analyzing time-dependent or evolving graphs is essential for applications requiring real-time monitoring, such as traffic flow analysis or financial forecasting. Advances in temporal pattern recognition within graphs enable models that capture dynamic behavior, supporting time-sensitive network analysis \cite{daverio2021temporal}.

\subsubsection{Graph Databases and Big Data Modeling}
The shift towards graph databases over relational databases is driven by the need for efficient storage and querying of complex relationships in large datasets. Graph databases, like Neo4j, are widely used in fraud detection, biological network analysis, and social media analytics due to their performance and flexibility \cite{paul2019review}.

\subsubsection{Graph Signal Processing}
Graph signal processing applies signal processing techniques to graph-structured data, enabling analysis on irregular domains like social networks or sensor arrays. This approach benefits fields like telecommunications, where efficient data representation and transmission are critical. Graph signal processing leverages graph-theoretical properties to enhance applications requiring high data integrity and efficiency \cite{dong2018graphsignal}.

\bibliography{refer.bib}

\end{document}